\documentclass[10pt,a4paper]{article}

\usepackage{amsmath,amsthm,amssymb,mathrsfs,bm,enumerate,physics}
\usepackage{tikz}
\usetikzlibrary{intersections,calc,arrows.meta}
\usepackage{url}
\usepackage{hyperref}
\usepackage{cleveref}
\usepackage{comment}
\newcommand\myshade{75}
\colorlet{mylinkcolor}{red}
\colorlet{mycitecolor}{green}
\colorlet{myurlcolor}{blue}
\hypersetup{colorlinks=true, anchorcolor=cyan, 
linkcolor=mylinkcolor!\myshade!black, 
citecolor=mycitecolor!\myshade!black,
urlcolor=myurlcolor!\myshade!black } 

\newcommand{\beq}{\begin{equation}}
\newcommand{\eeq}{\end{equation}}
\newcommand{\nn}{\nonumber}
\newcommand{\ra}{\rightarrow}
\newcommand{\lra}{\leftrightarrow}
\newcommand{\qqi}{q-q^{-1}}
\newcommand{\eps}{\epsilon}
\newcommand{\pqCom}[4]{ [\,{#1} \,, {#2}\,]_{({#3},{#4})}    }
\newcommand{\qCom}[3]{ [\,{#1} \,, {#2}\,]_{({#3})}    }

\newcommand{\Op}[3]{ \hat{#1}_{#2}^{({#3})}}

\newcommand{\Exp}[1]{\exp\left({#1}\right)}

\theoremstyle{definition}
\newtheorem*{rmk*}{$\ddagger$  Remark}

\theoremstyle{remark}
\newtheorem*{exm*}{$\spadesuit$ Example}

\makeatletter
\long\def\@makefntext#1{\parindent 1em\noindent
\@hangfrom{\hbox to 1.8em{\hss$^{\@thefnmark}$}}#1}
\makeatother
\begin{document}
\topmargin 0pt
\oddsidemargin 0mm
\renewcommand{\thefootnote}{\fnsymbol{footnote}}

\vspace*{0.5cm}


\begin{center}
{\Large Hom-Lie-Virasoro symmetries in Bloch electron systems \\
and quantum plane in tight binding models }
\vspace{1.5cm}

{\large Naruhiko Aizawa${\,}^{a,}$
\footnote{\,\,\, Corresponding author.  E-mail address: aizawa@omu.ac.jp}} 
\hspace{0.5cm}and \hspace{0.5cm}
{\large Haru-Tada Sato${\,}^{a,b}$}\\

{${}^{a}$\em Department of Physics, Graduate School of Science, \\Osaka Metropolitan University \\
Nakamozu Campus, Sakai, Osaka 599-8531, Japan}\\

{${}^{b}$\em Department of Data Science, i's Factory Corporation, Ltd.\\
     Kanda-nishiki-cho 2-7-6, Tokyo 101-0054, Japan}\\
%
\end{center}

\vspace{0.1cm}

\abstract{We discuss the Curtright-Zachos (CZ) deformation of the Virasoro algebra 
and its extentions in terms of magnetic translation (MT) group in a discrete Bloch 
electron system, so-called the tight binding model (TBM), as well as in its continuous system. We verify that the CZ generators are essentially composed of a specific 
combination of MT operators representing deformed and undeformed $U(1)$ 
translational groups, which determine phase factors for a $\ast$-bracket 
commutator. The phase factors can be formulated as a $\ast$-ordered product of the commutable $U(1)$ operators by interpreting the AB phase factor of discrete 
MT action as fluctuation parameter $q$ of a quantum plane. 
We also show that some sequences of TBM Hamiltonians 
are described by the CZ generators. 
}
\vspace{0.5cm}

\begin{description}
\item[Keywords:] quantum plane, $q$-deformation, Virasoro algebra, Hom-Lie algebra, magnetic translation, tight binding model 
\item[MSC:] 17B61, 17B68, 81R50, 81R60
\end{description}

%
%

\newpage
\setcounter{page}{1}
\setcounter{footnote}{0}
\setcounter{equation}{0}
\setcounter{secnumdepth}{4}
\renewcommand{\thepage}{\arabic{page}}
\renewcommand{\thefootnote}{\arabic{footnote})}
\renewcommand{\theequation}{\thesection.\arabic{equation}}

\section{Introduction}
\subsection{FFZ and Moyal deformation}
\indent

The Moyal sine algebra (FFZ algebra)~\cite{FFZ} is related to the Moyal bracket deformation, which is well-known as a Lie-algebraic deformation of the Poisson 
brackets and has been applied to describe noncommutative phenomena in various 
regions such as noncommutative geometry in string theories and quantum Hall physics.

The Moyal bracket and its star product are defined by~\cite{Moyal}
\begin{align}
& \{f(x,p),g(x,p)\}_\ast =\frac{2}{\hbar}\sin(\frac{\hbar}{2}\theta^{ab}\partial^a_1\partial^b_2)f(x,p)g(x,p)\,,\\
&f*g=\exp{i\frac{\hbar}{2}\theta^{ab} \partial_1^a \partial_2^b }f(x,p)g(x,p)\,,
\end{align}
where $\theta^{xp}=-\theta^{px}=-\theta$, and $\partial_1$ and $\partial_2$ denote 
forward (left) and backward (right) derivative operations respectively. 
The Moyal commutator is related to the Moyal bracket
\beq
[f,g]_\ast=f\ast g - g\ast f =i\hbar \{f,g\}_\ast
\eeq
and is often mentioned as Moyal deformation or Moyal quantization which can be 
regarded as a quantum deformation of the Dirac bracket
\beq
  i\hbar \{f,g\}_\ast \ra [f,g] \,.
\eeq
It is known that this quantization leads to the $SU(\infty)$ Lie algebra, so-called 
the Moyal sine algebra, if one takes 
the bases $\tau_{n,k}=e^{i(nx+kp)}$ of $T^2$ phase space~\cite{FFZ}:
\beq
[\tau_{n,k}, \tau_{m,l}] = 2i \sine(\frac{\hbar\theta}{2}(nl-mk)) \tau_{n+m,k+l} \,. \label{Msine}
\eeq
In the context of deformation quantization one may keep the star product structure
\beq
[\tau_{n,k}, \tau_{m,l}]_\ast = 2i \sine(\frac{\hbar\theta}{2}(nl-mk)) \tau_{n+m,k+l}\,. \label{ffz}
\eeq

What we call the FFZ algebra in this note is originally given by \eqref{Msine} with the introduction of  deformation parameter $q$ and the $q$-bracket for an arbitrary 
object $A$ 
\beq
q=e^{i\hbar\theta}\,,\quad  \mbox{and}\quad [A]=\frac{q^A -q^{-A}}{\qqi}\,. \label{qbraA}
\eeq
Changing the normalization
\beq
T_{(n,k)} =\frac{1}{\qqi} \tau_{n,k}
\eeq
we have the FFZ algebra
\beq
[T_{(n,k)},T_{(m,l)}]=[\frac{nl-mk}{2}] T_{(n+m,k+l)}\,, \label{FFZsine}
\eeq
where we assume that $\tau_{n,k}$ is generalized to an arbitrary operator behaving 
like the Moyal star products~\cite{FFZ} 
\beq
\tau_{n,k}\tau_{m,l} =q^{\frac{nl-mk}{2}} \tau_{n+m,k+l} \,. \label{FFZstar}
\eeq
It is well known that magnetic translation (MT) operators satisfy this 
fusion relation~\cite{Zaq}.

\subsubsection{noncommutative geometry}
\indent

Quantum field theory on noncommutative spaces~\cite{NCFT} is one of many 
applications of the Moyal brackets. The coordinates of the endpoints of open strings constrained to a D-brane in the presence of a constant Neveu-Schwarz B field are 
reported to be relevant to a noncommutative algebra\cite{SW}. The field theory on 
noncommutative boundary space (known as noncommutative field theory) has 
attracted much attention in string and M-theories, and the noncommutativity is 
expressed in the Moyal star bracket 
$[x^{\mu},x^{\nu}]_\ast=i\hbar\theta^{\mu\nu}\,.$ 
This noncommutative feature is understood to be originated from the same idea as 
noncommutative magnetic translations in two-dimensional quantum mechanics in a 
constant magnetic field~\cite{Zaq}. 

There is also an interesting topic in quantum gravity in a relevance to infinite 
dimensional symmetries. Moyal deformations of self-dual gravity have recently 
been studied in the context of noncommutativity  and $W_{1+\infty}$ 
algebra~\cite{BHS}, while the classical counterpart $w_{1+\infty}$ algebra~\cite{Winf} 
is related to soft graviton symmetries in asymptotically flat 4D quantum gravity~\cite{Strom}
\beq
[w_n^p,w_m^q]=(n(q-1)-m(p-1))w_{n+m}^{p+q-2}\,.  \label{winf1}
\eeq

\subsubsection{quantum Hall effect}
\indent

On the other hand, another type of $W_\infty$ algebra has been examined in 
quantum Hall physics \cite{QHE,GMP} and in the conformal field theory of edge 
excitations as well as in bulk physics extension~\cite{latestQH}. 
It is also known that $W_\infty$ is associated to the area-preserving 
diffeomorohisms (for a review see \cite{APD}) of incompressible fluids. 
The generators $\tilde{w}_n^k$ satisfy~\cite{CTZ}
\beq
[\tilde{w}_n^k, \tilde{w}_m^l]=((k+1)(m+1)-(n+1)(l+1))\tilde{w}_{n+m}^{k+l}\,, \label{winf2}
\eeq
and their generating functions $\rho(k,\bar{k})$ consisting of $\tilde{w}_n^k$ 
as Fourier components~\cite{CTZ2} obey the 
Girvin-MacDonald-Platzman algebra~\cite{GMP}
\beq
[\rho(k,\bar{k}),\rho(p,\bar{p})]=(e^{p\bar{k}/4}-e^{\bar{p}k/4})\rho(k+p, \bar{k}+\bar{p})\,.
\eeq
Interestingly, this relation can be identified with the FFZ commutation relation~\eqref{FFZsine} by changing the normalization
\beq
\rho(k,\bar{k})=e^{-\frac{k\bar{k}}{8}}W_{k,\bar{k}}\,,
\eeq
which leads to~\cite{CTZ2}
\beq
[W_{k,\bar{k}},W_{p,\bar{p}}]=2\sinh(\frac{p\bar{k}-\bar{p}k}{8})W_{k+p, \bar{k}+\bar{p}}\,.
\eeq

\subsection{quantm algebra and Hall effect} 
\indent

Another interesting aspect of quantum Hall effect is a certain relevance to 
integrable models in two-dimensional lattice systems. The algebraic approach of 
dynamical symmetries is useful when the Hamiltonian can be written in terms of 
the symmetry generators. The problem of Bloch electrons in a constant magnetic 
field can be solved by making use of a relation between the group of magnetic 
translations and $U_q(sl_2)$ quantum group~\cite{DJ} (strictly speaking the quantum 
algebra, which is a universal enveloping algebra with non-cocommutative Hopf algebra  structure~\cite{DJ,skr,frt,wor}). The Hamiltonian of a particle on a two-dimensional 
square lattice in the magnetic field (tight binding model) is composed of $U_q(sl_2)$ 
raising and lowering operators, which are expressed in $N$-dimensional Weyl bases 
$X$ and $Y$ with the commutation relation $qXY=YX$~\cite{WZ}. 
The same approach on a triangluar lattice is shown in~\cite{FK} adding the 
3rd basis $Z$ satisfying $qZX=XZ$ and $qYZ=ZY$. Spectra of these models  can be 
represented by means of solutions of Bethe ansatz type algebraic 
equations~\cite{WZ,FK,HKW}. The magnetic translations in the discretized systems 
are given by products of the Weyl base matrices, and their commutation 
relations are accompanied by a global phase factor such as $q^2T_x T_y=T_y T_x$ 
owing to the Weyl commutation relations. 

While in a continuous coordinate system, magnetic translations are made of 
differential operators, and their phase factors are no longer global but rather 
comprising local parameters as seen in \eqref{ffz}, and the Hamiltonian cannot be 
written in terms of those quantum group generators. Nevertheless it is interesting 
that there exist the generators of quantum algebra $U_q(sl_2)$ expressed by 
magnetic translations~\cite{uqsl2} in somewhat parallel form to the cases of the 
lattice systems. Furnished with the local parameters of Moyal type \eqref{ffz}, 
the generators can further be extended to $q$-deformed Virasoro 
(super)algebras~\cite{JS,KS} which are studied in field theory context~\cite{Chai,Belov,OPE1,OPE2,Noether,CP2}. 
Although these deformed algebras are not a quantum algebra but infinite 
dimensional Lie algebras, an interesting point is definitely an appearance of 
Moyal noncommutative feature of magnetic translations. 
Despite the clear correspondence of the tight binding Hamiltonian to the continuous 
system in a continuum limit, the inheritance of properties of magnetic translations 
such as quantum groups and the Weyl bases is still unclear. 
In other words, they are considered to disappear or not to be in the limit, 
regardless of having the similar characteristics mentioned above. 

%
%
Related to the quantum Hall effects and $q$-deformed algebras, there is also an interesting approach using Tsallis statistics~\cite{TS} 
to thermodynamic calculations~\cite{WSC,WSC2,Hassan}.

\subsection{quantum space and CZ algebra}
\indent

In order to reveal the truth and resolve such a conflicting situation, 
quantum groups and planes may help. Quantum groups~\cite{ma}  are 
another notion to describe noncommutative geometry, 
and they are formulated to be dual objects to quantum algebras~\cite{tak,maj}. 
They are deformations of matrix groups whose matrix entries obey certain 
commutation relations depending on a deformation parameter $q$. Quantum space  possesses noncommuting coordinates $X^i$ and their differentials 
$\partial_i = \dfrac{\partial}{\partial X^i}$  
which satisfy the relations
\beq
X^i X^j=B^{ij}_{kl} X^k X^l\,,\quad
\partial_i\partial_j=F^{lk}_{ji}\partial_k\partial_l \,,\quad
\partial_j X^i=\delta^i_j+C^{ik}_{jl} X^l \partial_k \,,
\eeq
where the coefficient matrices $B$, $C$ and $F$ satisfy certain Yang-Baxter relations~\cite{wz}. 
This ensures that the coordinates and their derivatives behave covariantly under the action of quantum group matrix. 
One of intriguing and the simplest example is the bosonic part of 
$GL_q(1,1)$ quantum superspace
\beq
\partial_x x=1+q^{-2}x\partial_x\,,
\eeq
since the Virasoro operators in this quantum space
\beq
L_n= - q^{-1}x^{n+1}\partial_x  \label{CZgl11}
\eeq
satisfy another version of $q$-deformed Virasoro algebra called the Curtright-Zachos (CZ) algebra~\cite{CZ} 
\beq
  \qCom{L_n}{L_m}{m-n} = [n-m] L_{n+m}\,,      \label{CZalg} 
\eeq
where the symbol $[x]$ is defined in \eqref{qbraA} and the bracket on LHS denotes 
the deformed commutator $[A,B]_{(x)}=q^x AB-q^{-x}BA\,.$ 
There are many results concerning representations of the CZ algebra such as:
$q$-harmonic oscillators~\cite{CKL}, 
central extensions~\cite{Cetal,AS} and operator product formula (OPE)~\cite{AS}, 
matrix representation~\cite{NQ}, quantum space diffrential calculi~\cite{superCZ,superCZ2,HHTZ}, 
and fractional spin representation~\cite{MZ}. 
Multi-parameter deformations~\cite{pqCZ,pqCZ2} and supersymmetric 
extensions~\cite{superCZ,superCZ2,super3} are also studied. 
Motivated by its application to physical systems, several investigations 
have been made: deformation of soliton equations~\cite{KdV,Poisson}, 
transformation to commutator form~\cite{Poisson}, 
Jacobi consistency conditions~\cite{AS,Pol}, and so on.

\subsection{Contents}
\indent

The purpose of this paper is three fold: first one is to clarify a dynamical  origin of the phase factors used in the commutator deformation of the algebra \eqref{CZalg}. 
Second is to clarify a relation between the phase factors and quantum plane picture 
which is relevant to physical models that possess a quantum algebra symmetry. 
Third is to present various ways of constructing the CZ generators. We investigate 
some properties of the CZ generators based on the algebras of MT and discrete 
magnetic translation (DMT), and discuss their relations to the Hamiltonian systems 
relevant to a tight binding model (TBM), which is a discrete model of a particle on a 
two-dimensional lattice in constant magnetic field. We expect to catch a glimpse of 
quantization of space, or rather the quantum plane structure, as a natural effect 
of space discretization in TBM, in addition to which is known to possess the 
quantum algebra symmetry $U_q(sl_2)$ (see Appendix~\ref{sec:TBM2CZ}).

In Section~\ref{sec:CZalg}, reviewing some properties of CZ algebra, we 
explain mathematical settings and MT algebras as our basic tool. 
The magnetic translations in TBM are reviewed in Section~\ref{CZ2TBM}. 
In Section~\ref{sec:CZ*}, we consider MT realizations of CZ algebras starting from 
a $q$-derivative representation of the algebra. We present a generalized algebra $CZ^\ast$, which includes $CZ$ algebra as a subalgebra of it. The generators of 
these algebras are composed of certain combinations of specific MT operators, 
and all these algebras as well as MT algebra can be encapsulated into a single 
expression of deformed commutators with a $\ast$-product structure.
\footnote{It is not clear that our $\ast$-product is the same as the Moyal 
$\ast$-product, however for convenience of terminology we adopt the same 
symbol without causing any serious confusion} 
The structure of $CZ^\ast$ generators consists of commutative MT and 
noncommutative MT parts. The commutative part plays the role of the 
fundamental $CZ$ algebraic relation (structure constant of the algebra), and the 
noncommutative part plays the role of a nonlocal operator of 
$\ast$-commutative translation (deformed $U(1)$) and the role of defining 
a weight for $\ast$-product phase factors as well. 

In Section~\ref{sec:NQCZ}, we construct a matrix representation of $CZ^\ast$ 
and verify the property of the representation given in Section~\ref{sec:CZ*} 
(we call the representation in Section~\ref{sec:CZ*}  ``commutative representation").
we investigate a mechanism how quantum plane structure and the $*$-bracket 
structure arise in $CZ^\ast$ algebra. $CZ^\ast$ contains two subalgebras $CZ^\pm$, 
and they correspond to two orthogonal directions on quantum plane in TBM. 
Section~\ref{sec:triviaCZ} outlines the role and significance of commutative 
representations. 
Section~\ref{sec:preTBM} defines the algebra family 
of $CZ^\ast$ representations that provide the TBM Hamiltonian sequences, 
in preliminary for the specific verification explained in Section~\ref{CZ2TBM}. 
Section~\ref{rmk:q-plane} presents another definition of our $\ast$-product. 
We first show that the commutative representations are composed of composite 
operators of DMT units in two directions. Then making use of the commutative 
representations and introducing a concept of $\ast$-ordering product, we 
formulate the appearance of the phase factor of the $CZ^\ast$ commutators 
on basis of the quantum plane picture of DMT in TBM. The derivation of the 
matrix representations of DMT is shown in Appendix~\ref{sec:Txy2XY}, 
where confirmation of their MT algebra is made as well. 

In Section~\ref{CZ2TBM}, we discuss the TBM Hamiltonian series universally 
represented by the matrix representations of $CZ^\ast$ algebras. 
In Section~\ref{sec:TBM}, we derive the DMT algebra (exchange, fusion and 
circulation rules) and observe the correspondence between the AB phase and 
the quantum plane fluctuations accompanied by the DMT movement. 
In Section~\ref{rmk:H_k1_model}, we describe the TBM Hamiltonian sequence 
$\check{H}_k$ using the $n=\pm1$ modes of $CZ^\ast$ family generators. 
Section~\ref{rmk:H_kn_model} discusses extensions to general modes of 
$CZ^\ast$ family in accord to the Hamiltonian systems $\hat{H}_n$ with 
the effective spacing of magnetic lattice extended from $1$ to $n$. 
In Section~\ref{rmk:H22}, more generic sequence $\hat{H}_{(n,k)}$ 
with the quantum plane fluctuation (power of $q$) extended from $1$ to $k$ 
is represented by the genuine $CZ^\ast$ generators.
As an aid to understanding some formulae used in Section~\ref{CZ2TBM}, 
we review the quantum group symmetry of the original TBM Hamiltonian, 
and add some remarks on $q$-inversion symmetry of the Hamiltonian in  Appendix~\ref{sec:TBM2CZ}.

\setcounter{equation}{0}
\section{Curtright-Zachos (CZ) algebra}
\label{sec:CZalg}
\indent

We make a brief review of mathematical settings on the CZ algebra. 
There are two types of bracket deformation, and they differ in how they introduce 
their phase factors. The way of introducing the phase factors  depends on explicit 
realizations of the generators of CZ algebra. Namely, there is no a priori way to 
determine how to introduce the phase factors and we have to start with a realization 
of the generators. However, the only known realization with physical implications is 
the deformation of harmonic oscillators, and we hence need other realizations to 
explore physical and mathematical properties of the CZ algebra. 
To this end, in Section \ref{sec:MT}, we introduce magnetic translation (MT) algebra, 
which is a typical realization of the Moyal sine algebra (FFZ algebra) in physical system. 
The property of generating phase factors when MT operators commute is compatible 
with the deformed commutation relation of the CZ algebra, and it is very convenient 
to investigate various features of the CZ algebra. 
Using MT representations in a later section, we will explain that there exists 
a larger algebra containing the CZ algebra, and investigate 
$\ast$-bracket formulation for these algebras altogether.

\subsection{bracket deformation}
\indent

The CZ algebra is neither a standard Lie algebra nor a quantum algebra, 
but a Hom-Lie algebra~\cite{Hom1,Hom2,Hom3}, which is a deformed Lie algebra 
satisfying skew symmetry and Hom-Jacobi conditions~\cite{Hom3}. 
The Hom-Lie algebras were originally introduced in~\cite{Hom1} as motivated by 
examples of deformed Lie algebras derived from twisted discretization of vector fields. 
The skew symmetry is provided by the relation
\beq
[L_n,L_m]_{(m-n)} =-[L_m,L_n]_{(n-m)}\,.
\eeq
The Hom-Jacobi condition of the CZ algebra is expressed as
\begin{equation}
(q^n + q^{-n}) [L_n, [L_m, L_{\ell}]_{(\ell-m)}]_{(m+\ell-n)} + \mathrm{cyc.\ perm.} =0\,,  \label{HLJ}
\end{equation}
and this condition consists of two parts. One is the Yang-Baxter associativity relation
\begin{equation}
[L_n, [L_m, L_{\ell}]_{(\ell-m)}]_{(m+\ell-2n)} + \mathrm{cyc.\ perm.} = 0\,, \label{YB}
\end{equation}
and the other is the consistency condition~\cite{AS}
\begin{equation}
[L_n, [L_m, L_{\ell}]_{(\ell-m)}]_{(m+\ell)} + \mathrm{cyc.\ perm.} = 0\,. \label{HL2}
\end{equation}
Representations in $q$-harmonic oscillator and quantum superspace as well as 
\eqref{CZgl11} are known to satisfy these conditions~\cite{superCZ,superCZ2,Poisson}.

Another feature that distinguishes it from the usual Virasoro algebra is 
the existence of the central element $S_0$~\cite{AS} 
\beq
S_0=1+(\qqi)L_0 \,,
\eeq
satisfying the commutation relations
\beq
S_0^{ k} L_n =q^{-2 nk}L_n S_0^{ k}\,,\quad
\mbox{or}\quad [S_0^{ k}, L_n]_{( nk,- nk)}=0\,. \label{center}
\eeq

Unlike the standard quantum and Lie algebras,  
the CZ algebra is defined in terms of the deformed bracket. 
We have two options to express the CZ algebra \eqref{CZalg} according to 
the way of defining deformed commutators: 
one is the exteria type
\begin{align}
    &\pqCom{A}{B}{x}{y}=q^xAB-q^yBA\,,  \label{com1}   \\
    &[A,B]_{(x)}=[A,B]_{(x,-x)} =q^x AB-q^{-x}BA\,,   \label{com2} 
\end{align}
and the other is the intrinsic type
\beq
     [A,B]_\ast=A*B-B*A\,.   \label{com3}
\eeq
The phase factors $q^x$ attached to the products $AB$ in Eqs.\eqref{com1} and \eqref{com2} are given by hand. 
On the other hand in \eqref{com3} the factors are implicitly included in the 
definiton of $A\ast B$, which is supposed to give rise to phase factors 
according to a certain product mechanism determined based on a phase space structure~\footnote{
We introduced the product as $(A_nB_m)_q=q^{m-n}A_nB_m$ in the original 
paper~\cite{AS}. } similarly to the Moyal commutators.
This does not necessarily mean to supply the Moyal star products. 

{}For the time being leaving the question open why the products are given in 
the following form and what its essential origin is, we suppose that
\begin{align}
&   L_n\ast L_m=q^{m-n}L_nL_m\,,    \label{LastL}  \\
&  S_0^k\ast L_n=q^{nk} S_0^k L_n\,,
\end{align}
and we then have 
\begin{eqnarray}
[L_n,L_m]_\ast &=& \qCom{L_n}{L_m}{m-n} \nn\\ 
                    &=& q^{m-n}L_nL_m - q^{n-m}L_mL_n \,.
\end{eqnarray}
In order to search for a certain origin of \eqref{LastL}, 
we later investigate MT representations of CZ algebras in Section~\ref{sec:CZ*}.
In Section \ref{rmk:q-plane}, we further explore mathematical features 
(quantum plane and $\ast$-ordered product) of discrete magnetic 
translations (DMTs) in tight binding model (TBM) on a two-dimensional 
square lattice as well.

\subsection{magnetic translation algebra (MTA)}
\label{sec:MT}
\indent

Magnetic translations $\hat{\tau}_{\bm{R}}$ of a charged particle in a 
constant magnetic field on a cotinuous coordinate surface are written in terms 
of differential operators in the following nonlocal way 
(choice of unit $\phi_0=hc/e=1$) 
\beq
\hat{\tau}_{\bm{R}}= e^{2\pi i\xi(\bm{x},\bm{R})} T_{\bm R}\,,
\eeq
where $T_{\bm R}$ is a translation by $\bm{R}$, and $\xi$ is a gauge fuction 
defined by
\beq
\bm{A}(\bm{x}+\bm{R})-\bm{A}(\bm{x})=\nabla \xi(\bm{x},\bm{R})\,. 
\eeq
They satisfy the following algebraic relations (exchange and fusion rules):
\beq
\hat{\tau}_{\bm R_1}\hat{\tau}_{\bm R_2}=
e^{ 2\pi i\phi }\hat{\tau}_{\bm R_2}\hat{\tau}_{\bm R_1} \,,
\eeq
\beq
\hat{\tau}_{\bm R_1}\hat{\tau}_{\bm R_2}=e^{ 2\pi i\xi(\bm{R_1},\bm{R_2}) }
    \hat{\tau}_{\bm{R_2}+\bm{R_1}}\,, 
\eeq
from which the circulation algebra is derived~\cite{Zaq}
\beq
\hat{\tau}_{\bm R_1}^{-1}\hat{\tau}_{\bm R_2}^{-1}\hat{\tau}_{\bm R_1}^{}\hat{\tau}_{\bm R_2}^{} =e^{2\pi i \phi}\,,
\eeq
where $\phi$ is a magnetic flux (proportional to the area 
$|\bm{R}_1\times \bm{R}_2|$ enclosed by the end points of circular operations). 
It is given by the differences of gauge function between two points
\beq
\phi=\xi(\bm{R_1},\bm{R_2})-\xi(\bm{R_2},\bm{R_1}) \,,
\eeq
where we denote
\beq
\bm{R_1}=(n_1,m_1)=(n,k)\,,\quad \bm{R_2}=(n_2,m_2)=(m,l)\,.
\eeq
In this paper we refer the set of exchange, fusion and circulation rules as 
MTA (magnetic translation algebra). 

Introducing the $q$ parameter with magnetic length $l_B=\sqrt{\hbar c/eB}$, 
unit length $a$ and magnetic field $B$, 
\beq
q=\Exp{2\pi i \frac{B a^2}{\phi_0}}=\Exp{i a^2 l_B^{-2}}\,,
\eeq
under the choice of symmetric gauge we have
\begin{align}
&\hat{\tau}_{\bm R_1}\hat{\tau}_{\bm R_2}=q^{nl-mk}\hat{\tau}_{\bm R_2}\hat{\tau}_{\bm R_1}\,,  \label{exchangeTR}  \\
&\hat{\tau}_{\bm R_1}\hat{\tau}_{\bm R_2}=q^{\frac{nl-mk}{2}}\hat{\tau}_{\bm{R_1}+\bm{R_2}}\,. \label{combineTR} \\
&\hat{\tau}_{\bm R_1}^{-1}\hat{\tau}_{\bm R_2}^{-1}\hat{\tau}_{\bm R_1}^{}\hat{\tau}_{\bm R_2}^{}=q^{nl-mk}\,. \label{circTR}
\end{align}
Applying the change of normalization to \eqref{combineTR}
\beq
\hat{\tau}_{\bm{R}_i} =\hat{\tau}_{n_i}^{(m_i)} = (\qqi) \Op{T}{n_i}{m_i}\,, \label{tau2T}
\eeq
we have the realization \eqref{FFZsine} of FFZ algebra~\cite{FFZ} by means of MTA
\beq
[\Op{T}{n}{k}\,, \Op{T}{m}{l}]=[\frac{nl-mk}{2}]\Op{T}{n+m}{l+k}  \,.  \label{eq:FFZ}
\eeq

As a simple example of $\hat{\tau}_n^{(k)}$ let us have a look at angular momentum 
phase space (together with a spin value $\Delta$)  on a unit circle 
$z=e^{i\theta}$ of  cylinder coordinate $w=\ln z$. Instead of the 
usual form $\hat{\tau}_{\bm{R}}=e^{\frac{i}{\hbar}\bm{R}\cdot{\bm \pi}}$ 
described by the gauge covariant derivatives $\hat{\pi}_i=\hat{p}_i+\frac{e}{c}A_i$, 
in this case we have
\beq
\Op{\tau}{n}{k} =\Exp{\frac{i}{\hbar}\bm{R}\cdot\bm{\Theta}} =
z^n q^{-k(z\partial +\frac{n}{2}+\Delta)} \label{taubyJ}\,,
\eeq
where
\beq
\bm{\Theta}=(\frac{\hbar}{l_B}\theta, -\frac{1}{l_B}\mathcal{J}_3 -\frac{\hbar}{l_B}\Delta)
=(-i\frac{\hbar}{l_B}\ln{z}, -\frac{\hbar}{l_B}(z\partial+\Delta))\,,
\eeq
\beq
\bm{R}=(nl_B,k\frac{a^2}{l_B})\,,\quad\quad 
\mathcal{J}_3=-i\hbar \partial_\theta = -i\hbar(x\partial_y-y\partial_x)\,.
\eeq
One can verify that \eqref{taubyJ} satisfies the MTA \eqref{exchangeTR}\eqref{combineTR}\eqref{circTR} (For simplicity, one may set $a=l_B=\hbar=1$.)

\setcounter{equation}{0}
\section{$\ast$-bracket formulation and $CZ^\ast$ algebra}
\label{sec:CZ*}
\indent

Starting from a $q$-differential operator expression of CZ generators, 
we present the realization of CZ algebra by MT operators, which leads to the idea of 
$\ast$-bracket formulation. In this section we discuss three types of CZ 
algebraic system $CZ^\pm$ and $CZ^\ast$. 

Section~\ref{sec:qdCZ} sets out the principle of giving rise to phase factors 
such that MT appears commutative (i.e., deformed U(1)). Setting weights on MT 
operators and CZ generators (special combinations of MT), we show that 
the $\ast$-brackets are realized by certain rules. Two closed subalgebras of MT, 
$\Op{T}{n}{0}$ and $\Op{T}{n}{2}$, play an important role in the construction 
of the CZ operators. The ordinary commutative operator $\Op{T}{n}{0}$ defines 
the structure constant of CZ algebra, and the other $\ast$-bracket commutative 
one $\Op{T}{n}{2}$ carries the weight of the CZ operators.

In Section~\ref{sec:CZpm*}, we show another special combination $CZ^-$, 
which transforms into $CZ^+$ mutually with $q$-inversion. Both $CZ^\pm$ can 
be combined into one algebraic system $CZ^\ast$ in the framework 
of $\ast$-brackets.

\subsection{$q$-derivative representation of $CZ$ algebra}
\label{sec:qdCZ}
\indent

Let us consider the $q$-analogue of differential operators $\partial_q$ 
(called the $q$-derivatives) defined by
\beq
  \partial_qf(z)=\frac{f(z)-f(zq^{-2})}{(\qqi)z} \,.  \label{delq}  
\eeq
This satisfies the following Leibniz rule and the formula
\beq
  \partial_q(f(z)g(z)) = g(z)\partial_qf(z)  + f(zq^{-2})\partial_q g(z)\,,  \label{delfg} 
\eeq
\beq  \partial_q z^n = q^{-n}[n] z^{n-1} \,. \label{delzn} \eeq
Defining the analogue of $l_n=-z^{n+1}\partial_z$ by replacing $\partial_z$ with 
$\partial_q$, 
\beq
\hat{L}_n=-z^{n+1}\partial_q   \label{Lnz}
\eeq
we can verify that $\hat{L}_n$ satisfy the CZ algebra eq.\eqref{CZalg}. 
\beq
  \qCom{\hat{L}_n}{\hat{L}_m}{m-n} = [n-m] \hat{L}_{n+m} \,.  \label{CZalg2} 
\eeq

Noticing eqs.\eqref{delq} and \eqref{Lnz}, the $q$-derivative can be rewritten by 
nonlocal expression of ordinary derivative $\partial_z$, and we have
\beq
\hat{L}_n = - z^{n}\frac{1-q^{-2z\partial}} {\qqi} \,.  \label{Ln_diff}
\eeq
This is related to the $q$-harmonic oscillators
\beq
a_q^\dagger=qz\,,\quad a_q=\frac{1}{qz}[z\partial]\,,\quad N=z\partial\,,
\eeq
which satisfy the following relations
\begin{align}
&a_qa_q^\dagger-qa_q^\dagger a_q=q^{-N}\,,\quad  [N]=a_q^\dagger a_q \\
&[N,a_q]=-a_q\,,\quad [N,a_q^\dagger]=a_q^\dagger\,, \\
&\hat{L}_n=-q^{-N}(a_q^\dagger)^{n+1}a_q\,. \label{Ln_qho}
\end{align}

Now, applying the magnetic translation \eqref{taubyJ} to \eqref{Ln_diff} 
with the notation \eqref{tau2T} we have another representation of 
the CZ generators 
\beq
  \hat{L}_n = -\Op{T}{n}{0} + q^{n+2\Delta} \Op{T}{n}{2}\,.  \label{CZT} 
\eeq
To make a contact with $*$-bracket formulation, we consider deformed commutators 
of $ \hat L_n $ and $ \hat T_m^{(k)}$. The basic algebraic relations \eqref{exchangeTR} and \eqref{combineTR} allow us to calculate various deformed commutators such as 
\begin{align}
&\pqCom{\hat{L}_n}{\Op{\tau}{m}{k}}{2m}{nk} = - q^{m+\frac{nk}{2}} [m] \Op{\tau}{n+m}{k} \,, \\
&\qCom{\hat{L}_n } {\Op{T}{m}{k} }{m-\frac{nk}{2} }  = -[m]\Op{T}{n+m}{k}\,,  \label{LT}  \\
&[\hat{L}_n,\Op{T}{m}{k}] = -[\frac{nk}{2}]\Op{T}{n+m}{k} -[m-\frac{nk}{2}]\Op{T}{n+m}{k+2}\, 
\end{align}
which are verified easily by utilizig the exchange relation between $\partial_q$ and 
$q^{-kz\partial}$
\beq
q^{-kz\partial}\partial_q = q^k \partial_q q^{-kz\partial} \,.
\eeq
At this stage, there is no guideline for choosing one out of them. 
We hence need a limiting condition or a principle to constrain the form of phase 
factors to be attached to the commutator deformations.  

One possible strategy is to follow the fact that 
magnetic translations should reduce to the usual translations when magnetic field 
vanishes. This naturally leads to the idea that $\Op{T}{n}{k}$ should satisfy a 
deformation of translational group $U(1)$. The most likely deformation is hence 
to attach phase factors to \eqref{com2} in a way to cancel the phase factors 
coming from the fusion rule \eqref{combineTR}, i.e., 
\beq
\qCom{\Op{T}{n}{k}}{  \Op{T}{m}{l}}{\frac{mk-nl}{2}} =0\,.  \label{TTalg}
\eeq
Let us refer the upper index $k$ in $\Op{T}{n}{k}$ to ``weight" and define the weight of $\hat{L}_n$ to be 2. Considering the following set of operators for all integers $n$ and $k$ 
\beq
{\mathscr M}=\{\; \hat{T}_n^{(k)}, \hat{L}_n  \;\}
\eeq
and allowing only the same phase factors as \eqref{TTalg} for $ \hat L_n$ 
as well as $ \hat T_n^{(k)},$ we thereby find the only three types of deformed 
commutators allowed
\beq
\qCom{\Op{T}{n}{k}}{  \Op{T}{m}{l}}{\frac{mk-nl}{2}}\,,\quad
\qCom{\hat{L}_n}{  \Op{T}{m}{l}}{m-\frac{nl}{2}}\,,\quad
\qCom{\hat{L}_n}{ \hat{L}_m}{m-n}\,,
\eeq
where we have put $k=2$ in the 2nd bracket reducing to \eqref{LT}, and $k=l=2$ 
in the 3rd one reducing to \eqref{CZalg2}. 
Note that we have omitted one type due to the skewness of the bracket \eqref{com2} 
\beq
\qCom{A}{B}{x}=-\qCom{B}{A}{-x}\,.
\eeq
In this way we adopt eqs.\eqref{CZalg2},\eqref{LT},\eqref{TTalg} for the 
fundamental set of closed algebras of $\mathscr{M}$, of which constituent 
\eqref{CZalg2} is the CZ algebra as a deformation of Virasoro algebra, and 
\eqref{TTalg} is a deformed $U(1)$ algebra of magnetic translation group as a 
deformation of $U(1)$ translation group.

In summary we confirm that $\ast$-bracket \eqref{com3} can be defined for every element $X_n^{(k)} \in \mathscr{M} $ of weight $k$  by
\beq
X_n^{(k)}\ast X_m^{(l)} = q^{-x} X_n^{(k)} X_m^{(l)}\,,\quad    x=\frac{nl-mk}{2}  \label{X*X}
\eeq
and we organize eqs.\eqref{CZalg2},\eqref{LT},\eqref{TTalg} into the single 
common bracket form
\begin{align}
&    [\Op{T}{n}{k},\Op{T}{m}{l}]_\ast =0\,,                 \label{*1}     \\
&   [\hat{L}_n, \hat{L}_m]_\ast = [n-m]\hat{L}_{n+m}\,,   \label{*2}   \\
&   [\hat{L}_n,\Op{T}{m}{l}]_\ast = -[m]\Op{T}{n+m}{l}\,.  \label{*3}  
\end{align}
Note that the algebra of central element \eqref{center} reads
\beq
[\hat{S}_0^k, \hat{L}_n]_\ast =0\,,\quad
\mbox{where}\quad \hat{S}_0=1+(q-q^{-1})\hat{L}_0\,,
\eeq
if we regard the weight of $\hat{S}_0^k$ as $2k$ in \eqref{X*X}. 

\subsection{$CZ^\pm$ and $CZ^\ast$ algebra}
\label{sec:CZpm*}
\indent

Inspecting the constitution of the CZ operators $\hat{L}_n$ in terms of subalgebras of $\Op{T}{n}{k}$, we consider $q$-inversion symmetry of $\mathscr{M}$, since 
the algebra set \eqref{*1}-\eqref{*3} is not invariant under the exchange of $q\leftrightarrow q^{-1}$.

In order to examine the structure of CZ algebra \eqref{*2}, 
let us set the same phase factor for $\Op{T}{n}{k}$ as for $\hat{L}_n$. 
Then we have
\beq
\qCom{\Op{T}{n}{k}}{\Op{T}{m}{l}}{m-n} = [\frac{n(l-2)-m(k-2)}{2}] \Op{T}{n+m}{k+l}\,. \label{Tmn}
\eeq
If we put $k,l=0,2$ in this equation, we obtain the following closed subalgebras 
on $\Op{T}{n}{k}$ with weight 0 and 2, which are the parts of $\hat{L}_n$ 
defined in \eqref{CZT}
\begin{align}
&\qCom{\Op{T}{n}{2}}{\Op{T}{m}{2}}{m-n}=0\,, \quad\hspace{40pt}   
  \qCom{\Op{T}{n}{0}}{\Op{T}{m}{0}}{m-n}=[m-n]\Op{T}{n+m}{0}\,, \label{T00} \\
&\qCom{\Op{T}{n}{2}}{\Op{T}{m}{0}}{m-n}= [-n]\Op{T}{n+m}{2}\,,  \quad\,\,
  \qCom{\Op{T}{n}{0}}{\Op{T}{m}{2}}{m-n}= [m]\Op{T}{n+m}{2}\,.    \label{T02}
\end{align}
According to these algebras, $\qCom{\hat{L}_n}{\hat{L}_m}{m-n}$ is verified 
to be closed without participations of any other weights:
\begin{eqnarray}
\qCom{\hat{L}_n}{\hat{L}_m}{m-n} &=& (q^n[n]-q^m[m])q^{2\Delta}\Op{T}{n+m}{2} 
+ [m-n]\Op{T}{n+m}{0}  \nn \\
&=&[n-m](q^{n+m+2\Delta}\Op{T}{n+m}{2}-\Op{T}{n+m}{0}) \nn \\
&=&[n-m]\hat{L}_{n+m}\,. \nn
\end{eqnarray}
We should note that $\Op{T}{n}{0}$ is a commuting local operator which does not 
include a differential operator w.r.t. $z$, while $\Op{T}{n}{2}$ is a noncommuting 
nonlocal differential operator. However this noncommutativity feature is reversed 
each other in the view of deformed commutator world as can be seen in eq.\eqref{T00}. 
If we find a set of subalgebras similar to eqs.\eqref{T00} and \eqref{T02}, 
it is possible to find another set of operators similar to $\hat{L}_n$.

It is in fact easy to find such a combination of $\Op{T}{n}{k}$ by considering 
$q$-inverted version of \eqref{Tmn}
\beq
\qCom{\Op{T}{n}{k}}{\Op{T}{m}{l}}{n-m} = [\frac{n(l+2)-m(k+2)}{2}] \Op{T}{n+m}{k+l}\,,  \label{Tmn-}
\eeq
and setting $k,l=0,-2$ we find another subalgebras corresponding to 
eqs.\eqref{T00} and \eqref{T02} 
\begin{align}
&\qCom{\Op{T}{n}{-2}}{\Op{T}{m}{-2}}{n-m}=0\,,\quad\hspace{20pt}    
  \qCom{\Op{T}{n}{0}}{\Op{T}{m}{0}}{n-m}=[n-m]\Op{T}{n+m}{0}\,,\label{T00-} \\
&\qCom{\Op{T}{n}{-2}}{\Op{T}{m}{0}}{n-m}= [n]\Op{T}{n+m}{-2}\,,\quad   
  \qCom{\Op{T}{n}{0}}{\Op{T}{m}{-2}}{m-n}= [-m]\Op{T}{n+m}{-2}\,.    \label{T02-}
\end{align}
Defining the $q$-inverted version of $\hat{L}_n$ with the new 
notation $\hat{L}_n^{\pm}$
\beq
\hat{L}_n^+ =\hat{L}_n\,, \quad\quad
\hat{L}_n^- =\Op{T}{n}{0} -q^{-n-2\Delta} \Op{T}{n}{-2}\,,   \label{Lpm1}
\eeq
we obtain 
\beq
\qCom{\hat{L}_n^+}{\hat{L}_m^+}{m-n} = [n-m]\hat{L}_{n+m}^+  \,,\quad\quad
\qCom{\hat{L}_n^-}{\hat{L}_m^-}{n-m} = [n-m]\hat{L}_{n+m}^- \label{newCZ1}
\eeq
\beq  \qCom{\hat{L}_n^\pm } {\Op{T}{m}{k} }{\pm m-\frac{nk}{2} }  
= -[m]\Op{T}{n+m}{k}\,,    \label{newCZ2}
\eeq
where we denote the two algebras in \eqref{newCZ1} as $CZ^\pm$ respectively. 
We notice that the signs of phase factors for $\hat{L}_n^\pm$ in \eqref{newCZ1} 
are completely opposite, while the phase factors in \eqref{newCZ2} are not symmetric w.r.t. the exchange of $\hat{L}_n^\pm$. This fact is originated in the $q$-inversion symmetry 
\beq
\Op{\hat{T}}{n}{k} \lra -\Op{T}{n}{-k}\,,\quad \hat{L}_n^+ \lra \hat{L}_n^-\,.  \label{qinv1}
\eeq
However, this clumsy combination perfectly disappears if we incorporate 
$\hat{L}_n^-$ into $\mathscr{M}$ assuming the weight of $\hat{L}_n^-$ 
to be $-2$. Thus changing its notation from $\mathscr{M}$ to 
\beq
{\mathscr M}^\ast =\{\; \hat{L}_n^\pm, \hat{T}_n^{(k) \;}  \}\,, \label{newM}
\eeq
we can make the new elements $\hat{L}_n^-$ participate in the following 
extended algebras including \eqref{*1}-\eqref{*3} with \eqref{X*X}:
\begin{align}
& [\hat{L}_n^\pm, \hat{L}_m^\pm]_\ast = [n-m]\hat{L}_{n+m}^\pm\,,  \label{*2pm} \\
&  [\hat{L}_n^\pm,\Op{T}{m}{l}]_\ast = -[m]\Op{T}{n+m}{l}\,.  \label{*3pm}  
\end{align}
Although the two algebras $CZ^\pm$ are expressed in \eqref{*2pm} respectively, 
the remaining intersecting algebra $[\hat{L}_n^+,\hat{L}_m^-]_\ast$ 
is yet open to incorporate into possible $\mathscr{M}^\ast$ algebra. 
Using \eqref{Lpm1} We then find 
\beq
\qCom{\hat{L}_n^+}{\hat{L}_m^-}{n+m}=q^{-m}[n]\hat{L}_{n+m}^+ - q^{n}[m]\hat{L}_{n+m}^-\,,   \label{minL+-2}
\eeq
and its phase factors are verified to certainly be given by \eqref{X*X} with 
the weights $k=\pm2$ applied to $X_n^{(k)}=\hat{L}_n^\pm$. 

Now let us show that \eqref{*2pm} and \eqref{minL+-2} can be organized into 
a closed algebra form. Introducing the notation $\eps,\eta$ to express 
the $\pm$ signs, \eqref{*2pm} and \eqref{minL+-2} read
\begin{align}
& [\hat{L}_n^\eps,\hat{L}_m^\eps]_\ast = [n-m] \hat{L}_{n+m}^\eps\,, \label{CZ*1}\\
& [\hat{L}_n^\eps,\hat{L}_m^\eta]_\ast =q^{\eta m}[n]\hat{L}_{n+m}^\eps - q^{\eps n}[m]\hat{L}_{n+m}^\eta\,. \quad (\eps\not=\eta) \label{CZ*2}
\end{align}
{}Furthermore using the formula
\beq
q^{\eps m}[n]-q^{\eps n}[m] =[n-m]\,,
\eeq
we can arrange the $CZ^\pm$ and their mixing algebras \eqref{CZ*1} and 
\eqref{CZ*2} in one final compact form
\beq
 [L_n^\eps,L_m^\eta]_\ast =q^{\eta m}[n]L_{n+m}^\eps - q^{\eps n}[m]L_{n+m}^\eta\,,  \label{CZCZ}
\eeq
which we shall denote the $CZ^\ast$ algebra.

In closing this section we put a brief remark on the $q$-derivative expression. 
\eqref{Lpm1} is equivalent to another $q$-derivative representation 
(similarly to the equivalence between \eqref{Ln_diff}  and \eqref{CZT})
\beq
\hat{L}_n^-=-z^{n+1} \partial^-_q =-z^n \frac{q^{2z\partial}-1}{\qqi}\,,
\eeq
where
\beq
\partial_q^-f(z) =\frac{f(zq^2)-f(z) }{z(q-q^{-1})} \,.
\eeq
Together with \eqref{Lnz}, we have a $q\lra q^{-1}$ symmetric form of the standard 
$q$-derivative (with $q$ substituted by $q^2$)
\beq
\partial_{q^2} :=\frac{\partial_q^+ + \partial_q^-}{q+q^{-1}} =
-\frac{\hat{L}^+_{-1}+\hat{L}_{-1}^-}{q+q^{-1}}=\frac{1}{z}[z\partial]_{q^2}\,.
\eeq
This type of $q$-derivative is used in operator product expansion form of $CZ^+$ algebra~\cite{AS,Poisson}.

\setcounter{equation}{0}
\section{Matrix representaions of $CZ^\ast$}
\label{sec:NQCZ}
\indent

This section studies matrix representations of $CZ^\ast$ and its related algebras. 
We discuss the role of $CZ^\pm$ commutative representations and their 
connection to quantum plane and $\ast$-bracket structure. We also present 
preliminary results about relationship of $CZ^\ast$ related algebras with the 
TBM Hamiltonian series. 

To begin with, we introduce the $CZ^\pm$ matrices in Section~\ref{sec:note}. 
The $CZ^\pm$ operator consists of a combination of commutative and 
non-commutative parts, as seen in the case of MT realization. 
Section~\ref{sec:triviaCZ} considers the meaning and role of commutative 
representations $X^n$ and $Y^n$, which play the same part as the MT operator 
$\Op{T}{n}{0}$. The commutative representation shows its significance in connection 
to quantum plane picture and the $\ast$-brackets. Concrete verification is done 
in Section~\ref{rmk:q-plane}. The matrix expression for $CZ^\ast$ is also given 
in Section~\ref{sec:triviaCZ}. Section~\ref{sec:preTBM} defines the representation sequence $CZ^\ast$ family, which is the sequence obtained by the replacement 
$q\ra q^k$. We introduce some matrix representations of $CZ^\ast$ family 
associated with the TBM Hamiltonian (concrete correspondence will be verified 
in Section~\ref{CZ2TBM}). 

In Section~\ref{rmk:q-plane}, considering a special combination of DMT that 
provides a commutative representation of $CZ^\pm$, we derive another definition 
of $\ast$-product based on the quantum plane picture of TBM. 
The AB phase (see Section~\ref{sec:TBM}) associated with the movement of 
particles by DMT corresponds to the fluctuation of the quantum plane, and the 
phase factor generated by the successive DMT operations is expressed by a 
certain ordered product to reproduce the $\ast$-bracket.

\subsection{cyclic representation of $CZ^{\pm}$ algebra}
\label{sec:note}
\indent

In order to examine the $CZ^\ast$ algebra, we must first set up its generators. 
Matrix representations of $CZ^-$ are already known~\cite{NQ}, but those of the 
other algebras are not. Thus we have to find a general expression for 
$CZ^+$ generators satisfying \eqref{*2pm}. The general expression obtained 
in this subsection does not always satisfy the $CZ^\ast$ relation \eqref{minL+-2}. 
This problem is solved by using an automorphism of $CZ^{\pm}$ in 
Section~\ref{sec:triviaCZ}. 

 Let us follow the basic idea given in~\cite{NQ}. The cyclic matrix representaion 
of $CZ$ generators are given if $q$ is a root of unity 
\beq
q=\Exp{\frac{2\pi i}{N}}\,,\quad N>2
\eeq
The minimal set of $L_n$ elements are as follows depending on whether $N$ is 
even or odd:
\begin{align}
& \{ L_{-m+1}, \cdots, L_{-1},L_0,L_1,\cdots,L_m\}  \quad \mbox{for}\quad N=2m \\
& \{ L_{-m}, \cdots, L_{-1},L_0,L_1,\cdots,L_m \} \quad \mbox{for}\quad N=2m+1\,.
\end{align}
The minimal set called the ``fundamental cell" \cite{NQ} and the translational group 
$G_N=\{ G_N^k: n\ra n+kN; k\in Z \}$ on a one-dimensional lattice of period $N$($>2$) 
gives rise to an algebra automorphism of $CZ$ algebra 
\beq
L_n \ra L_{n+kN}, \quad k=0,\pm1,\pm2,\cdots
\eeq
These correspond to a magnetic unit cell and the magnetic translation group in a Bloch 
electron system respectively.

Introducing the $N\times N $ Wyle base matricies $X$ and $Y$ satisfying $X^N=Y^N=1$
\beq
X=
\begin{pmatrix}
0 & 0 & 0 & \cdots\,0  & 1 \\
1 & 0 & 0 & \cdots\,0 & 0 \\
0 & 1 & 0 & \cdots\,0  & 0 \\
: & : & : &  & : \\
0 & 0 & 0 & \cdots\, 1 & 0\\
\end{pmatrix}\,,\quad
Y=
\begin{pmatrix}
q & 0 & 0 &   \cdots  & 0 \\
0 & q^2 & 0& \cdots  & 0 \\
0 & 0 & q^3 & \cdots & 0 \\
: &  :  & :    &  & : \\
0 & 0 & 0 & \cdots & q^N \\
\end{pmatrix}  \label{XYmat}
\eeq
or rather the component representation
\beq
(X)_{jk}=\delta_{j,k+1}\,,\quad (Y)_{jk}=q^j\delta_{jk}\,,\quad\mbox{for}\quad j,k \in [1,N] \quad(\mbox{mod}\,N)
\eeq
we find the cyclic matrix representation of $CZ^\pm$ algebra 
satisfying \eqref{*2pm} 
\beq
{\mathrm L}^\pm_n=\mp\left( \frac{1-{\mathrm Q}^{\pm2}}{q-q^{-1}}+A_n^{\pm} {\mathrm Q}^{\pm2} \right)
{\mathrm H}^n \,,\quad A_n^{\pm}=a_\pm+b(q^{\pm2n}-1)\,,    \label{QHCZ}
\eeq
where $a_{\pm}$ and $b$ are free parameters. ${\mathrm H}$ and ${\mathrm Q}$ are 
the matricies used in~\cite{NQ}. They are related to the Wyle base matricies through 
${\mathrm H}=X^{-1}=X^{T}$ and $q{\mathrm Q}=Y$, where $X$ and $Y$ satisfy the relation
\beq
Y^mX^n=q^{mn}X^nY^m\,.
\eeq

\subsection{roles of trivial form and $CZ^\pm$}
\label{sec:triviaCZ}
\indent

We here deal with ${\mathrm L}^\pm_n$ in parallel, in order to consider the extension 
of $CZ^\pm$ to $CZ^\ast$ later (see bottom of the subsection). As discussed later 
(Section~\ref{rmk:q-plane}) the phase factors of deformed commutators for 
${\mathrm L}^\pm_n$ correspond to translations accompanied by  phases $q^{\pm1}$ 
in two orthogonal directions on a quantum plane. This is one of the reasons why 
we extend the $CZ$ algebra to $CZ^\pm$ in this papar. 

Before discussing the $CZ^\ast$ extension, some remarks are in order. 
The first one is that the role of $Q^2$ as a central element~\cite{NQ} can be 
extended to $Q^{\pm2}$ for $CZ^\pm$ algebras as follows: 
\beq
[{\mathrm Q}^2, {\mathrm L}^+_n]_\ast=[{\mathrm Q}^2,{\mathrm L}^+_n]_{(n,-n)}=0\,,\quad
[{\mathrm Q}^{-2}, {\mathrm L}^-_n]_\ast=[{\mathrm Q}^{-2},{\mathrm L}^-_n]_{(-n,n)}=0\,,  \label{QL}
\eeq
where these elements are related to the central elements $S_0^\pm$ $(S_0^+=S_0)$
\beq
S_0^\pm=1\pm(\qqi)L_0^\pm = \{1-A_0^\pm (\qqi)\} {\mathrm Q}^{\pm2} \,.
\eeq
In the form of unified expression ($\eps=\pm$), one may have the relations 
\beq
[{\mathrm Q}^{\eps2k}, {\mathrm L}_n^\eps ]_{(\eps nk,-\eps nk)} =0\,,
\eeq
and
\beq
S_0^{\eps k} L_n^\eps =q^{-2\eps nk}L_n^\eps S_0^{\eps k}\,,\quad
\mbox{or}\quad [S_0^{\eps k}, L_n^\eps]_\ast
=[S_0^{\eps k}, L_n^\eps]_{(\eps nk,-\eps nk)}=0\,.
\eeq
{}From these, we again recognize $S_0^\pm \propto Q^{\pm2}$.  

The second remark is about trivial (commutative) representations
\beq
{\mathrm L}^{'\pm}_n=\frac{\mp g_n}{q-q^{-1}}\,,\quad   g_n g_m=g_{n+m}\,. \label{triviaL}
\eeq
Two examples of $g_n$ are given in~\cite{NQ}
\beq
g_n=c^n{\mathrm H}^n\,,\quad \mbox{or}\quad g_n=q^{cn^2}{\mathrm Q}^{2cn}{\mathrm H}^n\,. 
\quad\,\mbox{(for const. $c$)} \label{triviaX}
\eeq
As can be seen in \eqref{QHCZ}, ${\mathrm L}^{\pm}_n$ is a linear combination of trivial part ${\mathrm H}^n$ and nontrivial part ${\mathrm Q}^{\pm2}{\mathrm H}^n$. 
We now recall that $\hat{L}_n^\pm$ is also composed of trivial (commutative) 
$\Op{T}{n}{0}$ and nontrivial (noncommutative) $\Op{T}{n}{\pm2}$ parts as previously 
mentioned below \eqref{T00}. 
In this sense the commutative representation ${\mathrm H}^n$ (essentially $X^{-n}$) 
plays a key role in $CZ^\pm$ algebras. 
It is also interesting to note that the substitution
\beq
\Op{T}{n}{0}=\frac{q^n}{\qqi}{\mathrm H}^n\,,\quad 
\Op{T}{n}{\pm2}=\frac{q^{\pm2}}{\qqi}{\mathrm H}^n{\mathrm Q}^{\pm2}\,, \label{TTHQ}
\eeq
satisfy the same algebras as \eqref{T00}, \eqref{T02} \eqref{T00-} and \eqref{T02-}. 
Here we reverted the ordering of ${\mathrm Q}^{\pm2}{\mathrm H}^n$ 
due to \eqref{HLH}.

There is another trivial representation in terms of $Y$ 
\beq
g_n=c^n Y^n  \label{triviaY}
\eeq
and we further notice that 
$X^n$ and $Y^n$ play an important role 
to understand a relation between quantum plane and the $\ast$-bracket \eqref{com3}
as shall be discussed in Section \ref{rmk:q-plane}. There we show that the commutative representations are nontrivially realized by composite operators of DMT units in 
two directions on a magnetic lattice. Commuting operators again behave like 
noncommuting operators in the framework of quantum plane, which generates 
phase factors when operators are exchanged. It will turn out that the 
$\ast$-bracket for $X^n$ and $Y^n$ fit perfectly with this quantum plane picture 
in the system of TBM. 

To close this subsection, we present a matrix representation of $CZ^\ast$ algebra 
based on \eqref{QHCZ}. In order to complete the set of algebras \eqref{CZCZ}, 
eq.\eqref{minL+-2} the remaining algebra $[L^+_n,L^-_n]_*$ 
should be satisfied in addition to $CZ^\pm$ algebras \eqref{*2pm}. 

If we choose $b=0$ in the matrix repreesntation \eqref{QHCZ} denoting 
${\tilde {\mathscr L}}_n^\pm$, we have
\beq
{\tilde {\mathscr L}}_n^\pm=\mp \left( \frac{1- {\mathrm Q}^{\pm2}}{q-q^{-1}} + a_\pm{\mathrm Q}^{\pm2} \right){\mathrm H}^n\,, \label{tildeL}
\eeq
and this does not satisfy \eqref{minL+-2} but a slightly different one
\beq
\qCom{{\tilde {\mathscr L}}_n^+}{{\tilde {\mathscr L}}_m^-}{n+m}
=q^{m}[n]{\tilde {\mathscr L}}_{n+m}^+ 
- q^{-n}[m]{\tilde {\mathscr L}}_{n+m}^-\,. \label{tildeL+-}
\eeq
However, applying the following transformation that keeps $CZ^\pm$ unchanged 
\beq
{\mathscr L}_n^\pm ={\mathrm H}^n{\tilde {\mathscr L}}_n^\pm {\mathrm H}^{-n}\,,  \label{HLH}
\eeq
we verify that \eqref{minL+-2} is certainly satisfied, i.e.,
\beq
\qCom{{\mathscr L}_n^+}{{\mathscr L}_m^-}{n+m}
=q^{-m}[n]{\mathscr L}_{n+m}^+ - q^{n}[m]{\mathscr L}_{n+m}^-\,. \label{notildeL+-}
\eeq
Since $CZ^\pm$ is preserved under the transfomation \eqref{HLH}, 
${\mathscr L}_n^\pm$ also satisfy \eqref{*2pm} and \eqref{CZCZ}. 
Therefore the $CZ^\ast$ algebra is confirmed. 
Note that none of \eqref{tildeL+-} and \eqref{notildeL+-} holds for $b\not=0$, 
and we only consider the $b=0$ case hereafter.

\subsection{Preliminary representaions to TBM}
\label{sec:preTBM}
\indent

This is a preliminary section to discuss connections between $CZ^\ast$ algebra 
and TBM (tight binding model) Hamiltonain \eqref{MBH}. TBM is a two-dimensional lattice model which reproduces electron's Schr\"{o}dinger equation under static magnetic field 
in a continuum limit.
Our goal (see Section~\ref{CZ2TBM}) is to show that TBM Hamiltonians can be 
expressed in the $CZ^\ast$ generators ${\mathscr L}_{\pm1}^\pm$ following 
the basic idea presented in Appendix~\ref{sec:TBM2CZ}, however in order to get an 
overview at the moment we focus our attention to some representaions derived 
from \eqref{tildeL} and \eqref{HLH} prior to detailed investigation.

These representations are relevant to three methods of finding a relationship 
between $CZ^*$ algebra and TBM: 
(i) modification of Schr\"{o}dinger equations according to the structure of 
quantum planes, 
(ii) operator factorization, 
(iii) change of the parameter $q$ in $CZ^\ast$ algebra. 
Modifications of the $CZ^\ast$ are necessary in the latter two cases. 

The first representaion is given by 
\beq
{\mathscr L}_n^\pm=
\mp {\mathrm H}^n\frac{1\mp i q^{\pm 2}{\mathrm Q}^{\pm2}}{q-q^{-1}} 
= \mp X^{-n}\frac{1\mp i Y^{\pm2}}{q-q^{-1}} \,, \label{CZHQ2}
\eeq
with the choice
\beq
a_\pm=\frac{1\mp i q^{\pm 2}}{q-q^{-1}} \,.
\eeq
Since TBM Hamiltonian $\hat{H}$ is a linear combination in $X^{\pm1}$ and 
$Y^{\pm1}$ as seen in Appendix~\ref{sec:TBM2CZ}
\beq
\hat{H}(X,Y;q)=iY^{-1}(X^{-1}-X)+i((X^{-1}-X)Y\,. \label{H_XYq}
\eeq
$Y^{\pm 2}$ in this representation is a bit inconvenient. In this case we have to 
consider a slightly modified Hamiltonian in accordance with a different Schr\"{o}dinger equation on a quantum plane whose effective length and magnitude of phase fluctuation are doubled compared to the linear Hamiltonian system. (Section \ref{rmk:H22}). 

The second type of representation is for example 
\beq
{\mathscr L}_n^\pm
=\mp{\mathrm H}^n\frac{1-  q^{\pm 2}{\mathrm Q}^{\pm2}}{q-q^{-1}} 
= X^{-n}Y^{\pm1}[Z]\,,\quad Y=q^Z\,. \label{CZHZ}
\eeq
with the choice
\beq
a_\pm=\frac{1+ q^{\pm 2}}{q-q^{-1}}\,.
\eeq
In this case we can extract the linear TBM Hamiltonian by factoring out the 
operator $[Z]$ (for details see \eqref{factorHZ}). In order to see the factorization, 
we rather employ $CZ^{\ast}{}'$ 
the following algebra
\begin{align}
&\qCom{ L'^-_n}{ L'^-_m}{n-m}=[n-m] L'^-_{n+m}\,,  \label{TBCZ--}    \\
&\qCom{L_n^+}{{L'}_m^-}{n+m}=[n]L_{n+m}^+ - [m]{L'}_{n+m}^-\,,  \label{TBCZ+-}
\end{align}
by introducing 
\beq
L'^-_n=q^n L_n^-
\eeq
where arbitrary representations can be applied. 

The third type of candidates are not exactly the ${\mathscr L}_n^\pm$ but slightly modified operators ${\check{\mathscr L}}_n^\pm$, where we consider the algebra 
given by $Y^{\frac{1}{2}}$ instead of $Y$ in the first representation \eqref{CZHQ2};
\beq
{\check{\mathscr L}}_n^\pm
=\mp{\mathrm H}^n\frac{1\mp i q_1^{\pm 2}{\mathrm Q}_1^{\pm1}}{q_1^{}-q_1^{-1}} 
= \mp X^{-n}\frac{1\mp i Y_1^{\pm1}}{q_1^{}-q_1^{-1}} \,, \quad Y_1^2=Y \,. \label{CZHQ1}
\eeq
It is obtained by the replacement $q\ra q_1=q^\frac{1}{2}$ and by changing 
matricies $\{Y^2,{\mathrm Q}^2\}$ to $\{Y_1,{\mathrm Q}_1\}$ in  \eqref{CZHQ2}, 
where the condition $q^N=1$ should be changed to $q_1^N=1$ as well as matrix 
entries $q$ in $Y_1$ to $q_1$. Commutation relation $qXY=YX$ is then changed 
to $q_1XY_1=Y_1X$. After all, except for the change of power in $Y$ 
and ${\mathrm Q}$, everything is understood as $q\ra q_1=q^\frac{1}{2}$. 
{}For the sake of later conveniences, we introduce more general modification 
\beq
{\check{\mathscr L}}_n^\pm
= \mp X^{-n}\frac{1\mp i Y_k^{\pm1}}{q_k^{}-q_k^{-1}} \,, \label{CZHQk}
\eeq
\beq
Y_k=diag(q_k,q_k^2,\cdots,q_k^N)\,,\quad q_k^N=1\,,\quad q_kXY_k=Y_kX\,.
\eeq
The modified algebra of ${\check{\mathscr L}}_n^\pm$ is given by 
the replacement $q\ra q_k$ in $CZ^\ast(q)$, that leads to a sequence of algebras 
$CZ^\ast(q_k)$, and we refer to it as $CZ^\ast$ family (algebras). 
We then have
\beq
[{{\check{\mathscr L}}_n^\pm}, {{\check{\mathscr L}}_m^\pm}]_{\ast_k}
=[n-m]_k {\check{\mathscr L}}^\pm_{n+m}\,,  \label{TBHalg1}
\eeq
\beq
[{{\check{\mathscr L}}_n^+} ,{{\check{\mathscr L}}_m^-}]_{\ast_k}
=q_k^{-m}[n]_k{\check{\mathscr L}}_{n+m}^+ - q_k^{n}[m]_k{\check{\mathscr L}}_{n+m}^-\,, \label{TBHalg2}
\eeq
where
\beq
[n]_k:=\frac{q_k^n-q_k^{-n}}{q_k-q_k^{-1}}\,,
\eeq
\beq
[{{\check{\mathscr L}}_n^+}, {{\check{\mathscr L}}_m^-}]_{\ast_k}=
\left.\qCom{{\check{\mathscr L}}_n^+}{{\check{\mathscr L}}_m^-}{n+m}\right|_{q\ra q^k}\,.
\eeq
We assume $Y_k=Y^k$ and $q_k=q^k$ for $k\geq2$, and \eqref{CZHQ2} multiplied 
by $q+q^{-1}$ corresponds to the $k=2$ case. \eqref{CZHQ1} is regarded as an 
exceptional case $k=1$ with $Y_1^2=Y$ ($\lra q_1^2=q$). 
This is the outlined strategy of how to obtain the TBM form $\hat{H}(X,Y_k;q_k)$ 
from ${\check{\mathscr L}}_n^\pm$ by changing the $q$ parameter in $CZ^\ast$ representations. Details are explained in Section \ref{rmk:H_k1_model}.

\subsection{quantum plane and $\ast$-bracket}
\label{rmk:q-plane}
\indent

We discuss quantum plane picture of $CZ^\pm$ algebra in the framework of discrete 
magnetic translations $\hat{T}_x$ and $\hat{T}_y$ in TBM. 
Relations between the discrete magnetic translations (DMT) and the Wyle base matrices $X$ and $Y$ are summarized in Appendix \ref{sec:Txy2XY} for the convenience. 

Let us consider the matrix expressions of DMT, for example given by \eqref{TbyXY1}:
\begin{align} 
&\hat{T}_x=-iXY\,,\quad \hat{T}_y=-iY^{-1}X \\
&\hat{T}_x^\dagger=iY^{-1}X^{-1}\,,\quad \hat{T}_y^\dagger=iX^{-1}Y 
\end{align}
As shown in Figure~\ref{fig1}, these DMT operators describe the translations 
on a two-dimensional lattice $(m,n)$ in four directions, respectively.  

Consider the points A,B,C and D on the line $Y_j$ for $j=n+m$ fixed to a constant value, 
and a route of successive movements by DMTs from the point A.  
There are two shortest ways to get to B from A, namely via $Y_{j-1}$ and via $Y_{j+1}$. 
In the case of via $Y_{j-1}$, we have
\beq
 \hat{T}_y^\dagger \hat{T}_x=(X^{-1}Y)(XY)=q Y^2  \,,
\eeq
and we interpret this relation that $Y^2$ corresponds to a movement from 
A to B along $Y_j$ with a phase factor $q$ caused by the fluctuation via $Y_{j-1}$. 
If we go from A to C  via $Y_{j-1}$ twice, we understand that $Y^4$ is the moving 
operator along $Y_j$, and $q^2$ the fluctuation phase factor. 

Similarly in the case of via $Y_{j+1}$, we have
\beq
\hat{T}_x \hat{T}_y^\dagger=(XY)(X^{-1}Y)=q^{-1} Y^2\,,
\eeq
and interpret that $Y^2$ corresponds to a movement from 
A to B along $Y_j$ with a phase factor $q^{-1}$ caused by the fluctuation via $Y_{j+1}$. 
As to the movements from A to D, which is in a opposite direction to B along $Y_j$, 
we understand in a parallel way that $Y^{-2}$ corresponds to the movements from 
A to D with a phase factor $q^{\pm1}$ caused by the fluctuation via $Y_{j\pm1}$. 

Let us define a positive direction for $Y_j$ as the one with increasing $n$ of the 
vertical axis, and denote the numbers of fluctuations via $Y_{j+1}$ (resp. $Y_{j-1}$) 
by $k$ (resp. $l$) for positive direction along $Y_j$ (they are denoted by $-k$ and $-l$ 
for negative direction) when moving to an arbitrary point which is $k+l$ points 
away along $Y_j$. Then we can express the translation operator composed of $k+l$ 
translations along $Y_j$ in the following way with the total phase factor
\beq
q^{-k+l}Y^{2(k+l)}\,.    \label{Yphase}
\eeq
{}For example, in the case of fluctuating once via $Y_{j+1}$ and twice via $Y_{j-1}$ 
when moving to C from D in Figure~\ref{fig1}, it reads $qY^6$ since 
we have $k=1$ and $l=2$.

We now define the $\ast$-product as an ordered product as follows: 
(i) first put the operators with fluctuations toward a positive direction (i.e., via $Y_{j+1}$) 
in a leftward position, and those with negative fluctuations (via $Y_{j-1}$) 
in a rightward position. (ii) Second attach phase factors $q^{-1}$ for a positive 
fluctuation, and $q$ for negative one.

Then we can express \eqref{Yphase} using the $\ast$-product definition as
\beq
(Y^{2k} Y^{2l})_\ast=q^{l-k}Y^{2(k+l)} \,. \label{YYast}
\eeq
By the way, $Y^{2n}$ is nothing but the trivial representation \eqref{triviaL} with \eqref{triviaY}, and  we therefore have the $CZ^+$ algebra with the definition \eqref{YYast} 
\beq
[{\mathrm L}_n^{'+}, {\mathrm L}_m^{'+}]_\ast=[n-m]{\mathrm L}_{n+m}^{'+}\,,\quad 
{\mathrm L}_n^{'+}= \frac{-Y^{2n}}{q-q^{-1}}\,.  \label{CZ+Y}
\eeq
The commuting operator $Y^2$ on the line $Y_j$ acquires a nontrivial phase 
factor related to the $\ast$-product \eqref{YYast} as an effect of fluctuations 
via $Y_{j\pm1}$. The $\ast$-product plays the function of projecting a 
commuting operator product into a noncommuting one. 
We conclude that this fact is formulated by ${\mathrm L}^{'+}$, which is a trivial $CZ^+$ representation. In other words, we have obtained the picture that 
commuting translation operators on a quantum line $Y_j$ (one-dimensional 
quantum plane) raise phase factors as an effect of quantum fluctuation 
of the quantum plane.

\begin{figure}[htbp]
\begin{minipage}[b]{0.45\linewidth}
\centering
\begin{tikzpicture}
\draw[->,>=stealth,thin](-0.5,0)--(4.5,0)node[above]{$m$}; 
\draw[->,>=stealth,thin](0,-0.5)--(0,4.5)node[right]{$n$}; 
\draw[dotted](-0.5,1)--(4.3,1);\draw[dotted](-0.5,2)--(4.3,2);\draw[dotted](-0.5,3)--(4.3,3);\draw[dotted](-0.5,4)--(4.3,4);
\draw[dotted](1,0)--(1,4.3);\draw[dotted](2,0)--(2,4.3);\draw[dotted](3,0)--(3,4.3);\draw[dotted](4,0)--(4,4.3);
\draw[dotted](-0.5,4.5)--(4.5,-0.5)node[below]{$Y_{j+1}$}; 
\draw(-0.5,3.5)--(3.5,-0.5)node[below]{$Y_j$};     
\draw[dotted](-0.5,2.5)--(2.5,-0.5)node[below]{$Y_{j-1}$};  
\coordinate(D)at(3,0);\fill[black](D)circle(0.07);
\coordinate(A)at(2,1);\fill[black](A)circle(0.07);
\coordinate(B)at(1,2);\fill[black](B)circle(0.07);
\coordinate(C)at(0,3);\fill[black](C)circle(0.07);
\draw(2,1)node[below left]{A};
\draw(1,2)node[below left]{B};\draw(1,2)node[above right]{$Y^2$};
\draw(0,3)node[below left]{C};\draw(0,3)node[above right]{$Y^4$};
\draw(3,0)node[below left]{D};\draw(3,0)node[above right]{$Y^{-2}$};
\draw(2,2)node[above right]{$q^{-1}$};
\draw(1,3)node[above right]{$q^{-1}$};
\draw(3,1)node[above right]{$q$};
\draw(0,2)node[below left]{$q$};
\draw(1,1)node[below left]{$q$};
\draw(2,0)node[below left]{$q^{-1}$};
\draw[->,>=stealth,thick,dashed](2,1)--(2,0); \draw[->,>=stealth,thick,dashed](2,0)--(2.95,0);
\draw[->,>=stealth,thick,dashed](2,1)--(3,1); \draw[->,>=stealth,thick,dashed](3,1)--(3,0.05);
\draw[->,>=stealth,thick](2,1)--(1,1); \draw[->,>=stealth,thick](1,1)--(1,1.95);
\draw[->,>=stealth,thick](1,2)--(0,2); \draw[->,>=stealth,thick](0,2)--(0,2.95);
\draw[->,>=stealth,thick](2,1)--(2,2); \draw[->,>=stealth,thick](2,2)--(1.05,2);
\draw[->,>=stealth,thick](1,2)--(1,3); \draw[->,>=stealth,thick](1,3)--(0.05,3);
\draw[->,>=stealth,thick](3,3)--(4,3); \draw(3.5,3)node[below]{$\hat{T}_x^\dagger$};
\draw[->,>=stealth,thick](3,3)--(3,4); \draw(3,3.5)node[left]{$\hat{T}_y^\dagger$};
\draw[->,>=stealth,thick](4,4)--(4,3); \draw(4,3.5)node[right]{$\hat{T}_y$};
\draw[->,>=stealth,thick](4,4)--(3,4); \draw(3.5,4)node[above]{$\hat{T}_x$};
\end{tikzpicture}
\caption{Translations on $Y_j$ line ($j=m+n$)}
\label{fig1}
\end{minipage}
\begin{minipage}[b]{0.45\linewidth}
\centering
\begin{tikzpicture}
\draw[->,>=stealth,thin](-0.5,0)--(4.5,0)node[above]{$m$}; 
\draw[->,>=stealth,thin](0,-0.5)--(0,4.5)node[right]{$n$}; 
\draw[dotted](-0.5,1)--(4.3,1);\draw[dotted](-0.5,2)--(4.3,2);\draw[dotted](-0.5,3)--(4.3,3);\draw[dotted](-0.5,4)--(4.3,4);
\draw[dotted](1,0)--(1,4.3);\draw[dotted](2,0)--(2,4.3);\draw[dotted](3,0)--(3,4.3);\draw[dotted](4,0)--(4,4.3);
\draw[dotted](-0.5,2.5)--(2.5,-0.5);\draw(2,-0.25)node[below]{$j$}; 
\draw[dotted](-0.5,3.5)--(3.5,-0.5); \draw(2.75,0)node[below]{$j+1$};    
\draw[dotted](0,4)--(4,0);\draw(3.5,0.5)node[below]{$j+2$}; 
\draw[dotted](0.5,4.5)--(4.5,0.5);\draw(4.25,1)node[below]{$j+3$}; 
\draw[dotted](1.5,4.5)--(5,1);\draw(5,1.5)node[below]{$j+4$}; 
\draw[dotted](-0.5,0.5)--(3.5,4.5)node[left]{$X_{r+1}$}; 
\draw(-0.5,-0.5)--(4.1,4.1)node[above right]{$X_r$};     
\draw[dotted](0.5,-0.5)--(4.5,3.5)node[right]{$X_{r-1}$}; 
\coordinate(D)at(0,0);\fill[black](D)circle(0.07);
\coordinate(A)at(1,1);\fill[black](A)circle(0.07);
\coordinate(B)at(2,2);\fill[black](B)circle(0.07);
\coordinate(C)at(3,3);\fill[black](C)circle(0.07);
\draw(0,0)node[below right]{D};\draw(0,0)node[above left]{${\tilde X}^{2}$};
\draw(1,1)node[below right]{A};
\draw(2,2)node[below right]{B};\draw(2,2)node[above left]{${\tilde X}^{-2}$};
\draw(3,3)node[below right]{C};\draw(3,3)node[above left]{${\tilde X}^{-4}$};
\draw(1,2)node[above left]{$q$};
\draw(2,3)node[above left]{$q$};
\draw(2,1)node[below right]{$q^{-1}$};
\draw(3,2)node[below right]{$q^{-1}$};
\draw(0,1)node[above left]{$q^{-1}$};
\draw(1,0)node[below right]{$q$};
\draw[->,>=stealth,thick,dashed](1,1)--(0,1); \draw[->,>=stealth,thick,dashed](0,1)--(0,0.05);
\draw[->,>=stealth,thick,dashed](1,1)--(1,0); \draw[->,>=stealth,thick,dashed](1,0)--(0.05,0);
\draw[->,>=stealth,thick](1,1)--(2,1); \draw[->,>=stealth,thick](2,1)--(2,1.95);
\draw[->,>=stealth,thick](1,1)--(1,2); \draw[->,>=stealth,thick](1,2)--(1.95,2);
\draw[->,>=stealth,thick](2,2)--(3,2); \draw[->,>=stealth,thick](3,2)--(3,2.95);
\draw[->,>=stealth,thick](2,2)--(2,3); \draw[->,>=stealth,thick](2,3)--(2.95,3);
\end{tikzpicture}
\caption{Translations on $X_r$ line ($r=n-m$)}
\label{fig2}
\end{minipage}
\end{figure}

To complete the investigation, we have to consider another direction orthogonal to 
$Y_j$. The argument is straightforward, but attention should be paid to matrix 
normalization in order to parallel the discussion above. We thus elaborate on 
the details with reference to Figure~\ref{fig2}. 
Let us consider the points A,B,C and D on the line $X_r$ for $r=n-m$, and 
two-way successive movements by DMTs from A to B via $X_{r\pm 1}$. 
In the case of via $X_{r-1}$, we have
\beq
\hat{T}_y^\dagger \hat{T}_x^\dagger=-(X^{-1}Y)(Y^{-1}X^{-1})
=-X^{-2}=q^{-1} {\tilde X}^{-2}\,,\quad {\tilde X}=iXq^{-1/2}\,, \label{YXfusion}
\eeq
which means that ${\tilde X}^{-2}$ corresponds to a movement from A to B 
along $X_r$ with a phase factor $q^{-1}$ caused by the fluctuation via $X_{r-1}$. 
If we go from A to C via $X_{r-1}$ twice, we thus have the moving operator 
${\tilde X}^{-4}$ along $X_r$ giving rise to the factor $q^{-2}$ .  

Similarly in the case of via $X_{r+1}$, we have
\beq
\hat{T}_x^\dagger \hat{T}_y^\dagger=-(Y^{-1}X^{-1})(X^{-1}Y)
=-q^2 X^{-2}=q {\tilde X}^{-2}
\eeq
which means that ${\tilde X}^{-2}$ corresponds to a movement from 
A to B along $X_r$ with a phase factor $q$ caused by the fluctuation via $X_{r+1}$. 
The movements from A to D, which is in a opposite direction to B along $X_r$, are 
understood in a parallel way that ${\tilde X}^2$ corresponds to the movements 
from A to D with the factor $q^{\pm1}$ caused by the fluctuation via $X_{r\pm1}$. 

Let us define a positive direction for $X_r$ as the one with increasing $n$ on the 
vertical axis, and denote the numbers of fluctuations via 
$X_{r+1}$ (resp. $X_{r-1}$) by $k$ (resp. $l$) for positive directions 
along $X_r$ (they are denoted by $-k$ and $-l$ for negative directions) 
when moving to an arbitrary point which is $k+l$ points away along $X_r$. 
Then we can express the translation operator composed of $k+l$ translations 
along $X_r$ in the following way with the total phase factor
\beq
q^{k-l}{\tilde X}^{-2(k+l)} \,.   \label{Xphase}
\eeq
{}For example, in the case of fluctuating once via $X_{r+1}$ and twice via $X_{r-1}$ 
when moving to C from D in Figure~\ref{fig2}, it reads $q^{-1}{\tilde X}^{-6}$ since 
we have $k=1$ and $l=2$.

In the same way as $Y^2$, we here define the ordered product for ${\tilde X}$ as well: 
(i) first put the operators with fluctuations toward a positive direction (i.e., via $X_{r+1}$) 
in a leftward position, and those with negative fluctuations (via $X_{r-1}$) 
in a rightward position. (ii) Second attach phase factors $q$ for a positive fluctuation, 
and $q^{-1}$ for negative one. Note that the phases are inversed compared to 
the previous $Y^2$ case.

Then we can express \eqref{Xphase} using the $\ast$-product as
\beq
({\tilde X}^{-2k} {\tilde X}^{-2l})_\ast=q^{k-l}{\tilde X}^{-2(k+l)}\,. \label{XXast}
\eeq
Again, ${\tilde X}^{2n}$ is nothing but a trivial representation \eqref{triviaL} with \eqref{triviaX}, and we have the $CZ^-$ algebra with the definition \eqref{XXast} 
\beq
[{\mathrm L}_n^{'-}, {\mathrm L}_m^{'-}]_\ast=[n-m]{\mathrm L}_{n+m}^{'-}\,,\quad
{\mathrm L}_n^{'-}=\frac{{\tilde X}^{-2}}{q-q^{-1}} \label{CZ-X}
\eeq
We therefore verify that the same picture as $CZ^+$ holds. Namely  the 
commuting operator ${\tilde X}^2$ on the quantum line $X_r$ acquires a 
nontrivial phase factor related to the $\ast$-product \eqref{XXast} as an effect 
of quantum fluctuations via $X_{r\pm1}$. Since $X_r$ is orthogonal to $Y_j$, 
it can be said that ${\mathrm L}_n^{'\pm}$ are the algebras belonging to 
directions orthogonal to each other. 

We finally put a remark that ${\tilde X}^{-2}$ increases the position $j$ by 2 
along $X_r$, and effetive moving length of ${\tilde X}^{-1}$ may thus amount 
to $\Delta j=1$ if one applies a dual lattice. 
Similarly $Y^2$ increases $r$ by 2 along $Y_j$, and thus $Y^{-1}$ may 
effectively increase by $\Delta r=1$.

\setcounter{equation}{0}
\section{$CZ^\ast$ and TBM Hamiltonians}
\label{CZ2TBM}
\indent

In Section~\ref{CZ2TBM}, we show that the matrix representation of TBM corresponds to the Wyle representation of $CZ^\ast$, which describes the Hamiltonian sequence covering various magnetic lattices.
In Section~\ref{sec:TBM}, deriving the DMT algebra (exchange, fusion and 
circulation rules) in TBM, we comment on its relation to the quantum plane picture. 
In Section~\ref{rmk:H_k1_model}, we show that the TBM Hamiltonian sequence 
can be described using the $\pm1$ modes of the matrix representation of 
the $CZ^\ast$ algebra family. 
Section~\ref{rmk:H_kn_model} discusses extensions to general modes. 
The power of $X$ corresponds to the Hamiltonian with the effective spacing 
of the magnetic lattice expanded from $1$ to $n$. The power of $Y$ corresponds 
to the Hamiltonian sequence that extends the quantum plane fluctuation 
(order of $q$) from $1$ to $k$ (Section~\ref{rmk:H22}). 
These Hamiltonians can be represented by the $CZ^\ast$ generators. 

\subsection{DMT and quantum plane inTBM}
\label{sec:TBM}
\indent

The purpose of this subsection is to verify the quantum plane picture of tight 
binding model (TBM) by showing that discrete magnetic translations (DMTs) satisfy 
the same properties as the MTA (exchange, fusion and circulation) of the continuous magnetic translations reviewed in Section~\ref{sec:MT}. In contrast to the 
discussion in Section~\ref{rmk:q-plane}, we do not use the matrix representation 
of DMT. As a result of this picture, TBM can be regarded as a Hamiltonian 
system constructed on a quantum plane. 

TBM Hamiltonian is given by DMT on a two-dimensioanl lattice as follows~\cite{WZ,FK,HKW,HH}:
\beq
H=\hat{T}_x+\hat{T}_y+\hat{T}^\dagger_x+\hat{T}^\dagger_y\,, \label{TBH}
\eeq
\beq
\hat{T}_x=\sum_{n,m}e^{i\theta_{mn}^x}c^\dagger_{m+1,n}c_{m,n}\,,\quad 
\hat{T}_y=\sum_{n,m}e^{i\theta_{mn}^y}c^\dagger_{m,n+1}c_{m,n}\,.    \label{TBTxTy}
\eeq
where $c^\dagger_{m,n}$ and $c_{m,n}$ represent the creation/annihilation 
operators at each site of $(m,n)$, and $\theta^x_{mn}$ and $\theta^y_{mn}$ are 
the $AB$ phase associated with the unit movement length $a$ in each direction 
of $x$ and $y$
\beq
\theta^x_{mn}: (m,n)\ra (m+1,n)\,,\quad \theta^y_{mn}: (m,n)\ra(m,n+1)\,.
\eeq
Introducing the wave function
\beq
\Psi=\sum_{m,n}\psi_{m,n}c^\dagger_{m,n}\ket{0}=\sum_{m,n}\psi_{m,n}\ket{\psi_{m,n}}\,,\quad \psi_{m,n} \in\mathbf{C}\,,
\label{wavef}
\eeq
the eigenvalue equation $H\Psi=E\Psi$ with \eqref{TBH} is known to reduce to the following Schr\"{o}dinger equation~\cite{WZ,HKW}
\beq
e^{i\theta^x_{m-1,n}}\psi_{m-1,n}+e^{i\theta^y_{m,n-1}}\psi_{m,n-1}+e^{-i\theta^x_{m,n}}\psi_{m+1,n}
+e^{-i\theta^y_{m,n}}\psi_{m,n+1}=E\psi_{m,n}\,. \label{TBSeq}
\eeq
Note that this reproduces the continuous Schr\"{o}dinger equation for the 
Hamiltonian $H'=-tH$ at the order of $\mathcal{O}(a^2)$,
\beq
\frac{1}{2m}(\bm{p}+\frac{e}{c}\bm{A})^2\psi_{m,n}=\mathcal{E}\psi_{m,n}\,,\quad \mathcal{E}=\frac{\hbar^2}{2ma^2}\frac{E+4t}{t} \,,
\eeq
where $\mathcal{O}(a)$ vanishes in the continuum limit $a\ra0$.

Having the formulae from \eqref{TBTxTy} and \eqref{wavef}, 
\begin{align}
&\hat{T}_x\hat{T}_y\ket{\psi_{m,n}}=e^{i\theta^x_{m,n+1}+i\theta^y_{m,n}}\ket{\psi_{m+1,n+1}}\,,  \label{TxTypsi}\\
&\hat{T}_y\hat{T}_x\ket{\psi_{m,n}}=e^{i\theta^y_{m+1,n}+i\theta^x_{m,n}}\ket{\psi_{m+1,n+1}}\,, \label{TyTxpsi}
\end{align}
we obtain the exchange and circulation algebras
\begin{align}
&\hat{T}_y\hat{T}_x\ket{\psi_{m,n}}=e^{2\pi i\phi}\hat{T}_x\hat{T}_y\ket{\psi_{m,n}}\,, \label{TxTyCom}\\
&\hat{T}_y^\dagger\hat{T}_x^\dagger\hat{T}_y\hat{T}_x\ket{\psi_{m,n}}=e^{2\pi i\phi}\ket{\psi_{m,n}}\,, \label{TxTyCirc}
\end{align}
where
\beq
2\pi\phi=(\theta^y_{m+1,n}-\theta^y_{m,n})-(\theta^x_{m,n+1}-\theta^x_{m,n})\,. \label{phaseTBM}
\eeq

Concerning the fusion algebra, we have to define new composite operator 
$\hat{T}_{x+y}$ satisfying the following fusion relations with phase factor $\xi_{m,n}$, 
which will be determined later
\begin{align}
&\hat{T}_x\hat{T}_y\ket{\psi_{m,n}}:=e^{i\xi_{m,n}}\hat{T}_{x+y}\ket{\psi_{m,n}}\,,\\
&\hat{T}_y\hat{T}_x\ket{\psi_{m,n}}:=e^{-i\xi_{m,n}}\hat{T}_{x+y}\ket{\psi_{m,n}}\,.
\end{align}
Together \eqref{TxTypsi}, \eqref{TyTxpsi} and these, we have
\begin{eqnarray}
\hat{T}_{x+y}\ket{\psi_{m,n}}&=&e^{-i\xi_{m,n}}e^{i\theta^x_{m,n+1}+i\theta^y_{m,n}}\ket{\psi_{m+1,n+1}} \\
&=&e^{i\xi_{m,n}}e^{i\theta^y_{m+1,n}+i\theta^x_{m,n}}\ket{\psi_{m+1,n+1}} \,.
\end{eqnarray}
Using \eqref{phaseTBM}, the consistency condition of the r.h.s of this equation reads
\beq
2\pi\phi+2\xi_{m,n}=0\,,
\eeq
and therefore we have the fusion algebra
\begin{align}
&\hat{T}_x\hat{T}_y\ket{\psi_{m,n}}:=e^{-i\pi\phi}\hat{T}_{x+y}\ket{\psi_{m,n}}\,,
\label{DMTfusion} \\
&\hat{T}_y\hat{T}_x\ket{\psi_{m,n}}:=e^{i\pi\phi}\hat{T}_{x+y}\ket{\psi_{m,n}}\,.
\end{align}

Introducing the parameter $q$ as
\beq
q=e^{-i\pi\phi} \,,  \label{localq}
\eeq
we summarize the exchange, fusion and circulation as follows
\beq
\hat{T}_x\hat{T}_y = q^2\hat{T}_y\hat{T}_x\,,\quad
\hat{T}_x\hat{T}_y=q\hat{T}_{x+y}\,,\quad
\hat{T}_y^\dagger\hat{T}_x^\dagger\hat{T}_y\hat{T}_x =q^{-2}\,.\label{DMTalg}
\eeq
If we combine the fusion and exchange rules into
\beq
\hat{T}_{x+y}=q^{-1}\hat{T}_x\hat{T}_y = q\hat{T}_y\hat{T}_x \,,
\eeq
this equation can be interpreted as follows: the translation $\hat{T}_{x+y}$ 
corresponds to the one along the line $X_r$ defined in Section~\ref{rmk:q-plane}, 
and the first and second equalities imply mutually different phase factors 
$q^{\pm1}$ in accordance with the fluctuations taking detours to adjacent 
$X_{r\pm1}$ and back to $X_r$, corresponding to different ordering of $\hat{T}_x$ 
and $\hat{T}_y$ operators. 
Note that the phase factor $q$ given by \eqref{localq} is not necessarily a 
constant because $\phi$, given by \eqref{phaseTBM},  depends on its site $(m,n)$. 
As discussed in Section~\ref{rmk:q-plane}, the $q$ can be regarded as the 
fluctuation of quantum plane, and it is related to the AB phases as seen 
in \eqref{phaseTBM}. 

This interpretation suggests that the $\ast$-products are generated by 
the quantization of space (quantum plane) in view of discretization. 
It is interesting that the quantum plane picture can be understood as 
the underlying structure before a periodic condition is taken into account. 

\subsection{$CZ^\ast$ and TBM Hamiltonian family}
\label{rmk:H_k1_model}
\indent

Hereafter we discuss the relations between TBM Hamiltonian $\hat{H}$ and 
the $CZ^\ast$ matrix representations ${\mathscr L}^\pm_{\pm1}$ defined in 
\eqref{CZHQ2}-\eqref{TBCZ--}. 
Since our three representations \eqref{CZHQ2},\eqref{CZHZ} and \eqref{CZHQ1} 
have different $Y$-dependence from $\hat{H}$ (see \eqref{MBH}), 
it is not straightforward to find their relationships. 
For convenience of discussion, we explicitly show the dependence on deformation parameters $q$ and matrix sizes $N$ in $CZ^\ast$ and TBM Hamiltonian $\hat{H}$, 
such as $CZ^\ast(q,N)$ and $\hat{H}(q,N')$, where the latter TBM matrix sizes are 
given by $N'=2Q$. 
Although the same symbols for $q$ and $N$ are employed in both $CZ^\ast$ and 
$\hat{H}$, they are originally introduced independently, and hence generally different. 
Then denoting  the deformation parameter of TBM by $q_k$ when both $q$ are 
related, we consider the situation that the Hamiltonian $\hat{H}(q_k,Q)$ coincides 
with $\check{H}_k$ which is a linear combination of $CZ^\ast(q_k,N)$ generators. 
Keeping the relation of $CZ(q_k,N)$ to its parent $CZ(q,N)$, and providing 
a certain relation between $N$ and $Q$, we are going to determine the values of 
$q$ and $q_k$ in each case. (In the case of the factorization \eqref{CZHZ}, 
we do not have to consider this issue, because $q_k$ coincides with $q$, 
which is nothing but a matrix element of $Y$ of size $N=2Q$.) 

Let us first consider the second type representation, that is the factorized form 
\eqref{CZHZ}, where the powers of $Y$ coincide with those in $\hat{H}$ up to 
the factorization of $[Z]$. In this case the $CZ^\ast$ algebra is slightly modified 
to the $CZ^{\ast'}$ algebra, which is given by ${\mathscr L}_n^+$ and 
${\mathscr L}_1^{'-}$
\beq
{\mathscr L}_n^{'-}=q^n{\mathscr L}_n^-\,,
\eeq
satisfying  \eqref{*2pm}, \eqref{TBCZ--}, \eqref{TBCZ+-}. 
It is convenient to define $\hat{H}_Z$ by using the $CZ^{\ast'}$ generators as
\beq
\hat{H}_Z=i({\mathscr L}_1^+-{\mathscr L}_{-1}^+)+i({\mathscr L}_1^{'-}-{\mathscr L}_{-1}^{'-})\,.
\eeq
Using the relation $qXY=YX$, we find that $[Z]$ is factorized from $\hat{H}_Z$ as
\beq
\hat{H}_Z=i(X^{-1}-X)Y[Z]+iY^{-1}(X^{-1}-X)[Z] =\hat{H}[Z]   \label{factorHZ}
\eeq
and the TBM Hamiltonian \eqref{MBH} is therefore related to the 
$CZ^\ast$ generators in the following form
\beq
\hat{H}=\hat{H}_Z [Z]^{-1}\,.
\eeq
This means that eigenvalues of $\hat{H}_Z$ are given by the product of 
the TBM eigenvalues \eqref{TBMwave} and the phase factor matrix $[Z]$,
\beq
\hat{H}_Z\tilde{\psi}=E[Z]\tilde{\psi}\,,\quad  \tilde{\psi}=[Z]^{-1}\psi\,.
\eeq

Next, let us consider the case \eqref{CZHQ1}, which is one of the third types and 
its corresponding algebra is $CZ^\ast(q_1)$ defined in \eqref{TBHalg1} and \eqref{TBHalg2}. 
We need the notation of the Hermite conjugation for $\check{{\mathscr L}}_n^\pm$
\beq
\check{{\mathscr L'}}_{n}^\pm=\mathrm{H}^{-n}\check{{\mathscr L}}_{n}^\pm
\mathrm{H}^n\,,\quad\quad
(\check{{\mathscr L}}_n^\pm)^\dagger=\check{{\mathscr L'}}_{-n}^\mp\,,
\eeq
where $\check{{\mathscr L'}}_{n}^\pm$ satisfy the same algebra of $\check{{\mathscr L}}_{n}^\pm$. This relation is completely the same as that of ${\mathscr L}_n^\pm$ and 
$\tilde{{\mathscr L}}_n^\pm$ shown in \eqref{HLH}:
\beq
\tilde{{\mathscr L}}_{n}^\pm=\mathrm{H}^{-n}{\mathscr L}_{n}^\pm
\mathrm{H}^n\,,\quad\quad
({\mathscr L}_n^\pm)^\dagger=\tilde{{\mathscr L}}_{-n}^\mp\,.
\eeq
Defining $\check{H}_1$ in terms of the $CZ^\ast(q_1)$ generators 
$\check{{\mathscr L}}_{\pm1}^+$ and $(\check{{\mathscr L}}_{\pm1}^+)^\dagger$
\beq
\check{H}_1=(q_1-q_1^{-1})(\check{{\mathscr L}}_1^+ -\check{{\mathscr L}}_{-1}^+)
  +(q_1-q_1^{-1})(\check{{\mathscr L'}}_1^- -\check{{\mathscr L'}}_{-1}^-) \,,
\eeq
and substituting \eqref{CZHQ1} on the r.h.s. of this, we obtain the same form as 
the TBM Hamiltonian $\hat{H}$
\beq
\check{H}_1=i(X^{-1}-X)Y_1+iY_1^{-1}(X^{-1}-X) =\hat{H}(X,Y_1; q_1)\,, \label{checkH1}
\eeq
where $Y_1$ and $q_1$ are substituted for $Y$ and $q$ in \eqref{H_XYq}.
In this representation we have the conditions $X^N=Y_1^N=1$ as well as $Y^N=1$. 
Both should be satisfied, and it is realized in the following way: 
$Y_1$ with the relation $q=q_1^2$ is related to $Y$ set in the $CZ^\ast$ representation \eqref{CZHQ2} by the relation $Y_1^2=Y$. Notice that it does not mean that  \eqref{CZHQ2} coincides with \eqref{CZHQ1}. If we choose $N=2Q$ 
remembering that $2Q$ is the matrix size of $\hat{H}$, we have
\beq
Y_1^{2Q}=1\,,\quad q_1^{2Q}=1\,,\quad\therefore\,q_1=e^{\pm i\pi\phi}\,,
\eeq
as well as for $Y$ and $q$
\beq
Y^Q=1\,,\quad q^Q=1\,,\quad \therefore\, q=e^{\pm 2\pi i\phi}\,,
\eeq
where the double sign $\pm$ is introduced for a complex conjugation system.
Thus we have
\beq
(Y_1)_{jk}=q_1^j\delta_{jk}\,,\quad (Y)_{jk}=q^j\delta_{jk}\,,\quad
Y_1^{N}=Y^{\frac{N}{2}}=1\,.  \label{matY1}
\eeq
The matrix representation \eqref{CZHQ1} describes the TBM Hamiltonian 
$\check{H}_1$ given by \eqref{checkH1} not with $q=e^{-i\pi\phi}$ but with 
\beq
q=e^{\pm 2\pi i\phi}\,,\quad\mbox{or}\quad q_1=e^{\pm i\pi\phi} \,.
\eeq
This adjustment corresponds to the manipulation to replace $q$ by $q_1$ in 
\eqref{MBeq} adopting $q=e^{- 2\pi i\phi}$, instead of using parametrization 
\eqref{q-phi} when driving the Schr\"{o}dinger equation \eqref{MBeq}. 

{}Finally we deal with the rest of all, the first type \eqref{CZHQ2} and the third type 
\eqref{CZHQk} in the same formalism $CZ^\ast(q_k)$, since \eqref{CZHQ2} is 
a special case of the third type with $k=2$. Defining $\check{H}_k$ as
\beq
\check{H}_k=(q_k-q_k^{-1})(\check{{\mathscr L}}_1^+ -\check{{\mathscr L}}_{-1}^+)
  +(q_k-q_k^{-1})(\check{{\mathscr L'}}_1^- -\check{{\mathscr L'}}_{-1}^-) \,, \label{TBH_k}
\eeq
and substituting \eqref{CZHQk} on the r.h.s., we obtain for $k=2$
\beq
\check{H}_2=i(X^{-1}-X)Y_2+iY_2^{-1}(X^{-1}-X) =\hat{H}(X,Y_2; q_2)\,,
\eeq
where $q_2=q^2$. In this representation we have the condition $X^N=Y_2^N=1$. 
Note that $Y^N\not=1$ this time. $Y_2$ with the relation $q_2=q^2$ is related 
to $Y$ in the $CZ^\ast$ representation \eqref{CZHQ2} by the relation $Y_2=Y^2$. 
If we choose $N=Q$, we have
\beq
Y_2^{Q}=1\,,\quad q_2^{Q}=1\,,\quad\therefore\,q_2=e^{\pm 2\pi i \phi}\,,
\eeq
as well as for $Y$
\beq
Y^{2Q}=1\,,\quad q^{2Q}=1\,,\quad \therefore\, q=e^{\pm \pi i\phi}\,.
\eeq
Thus we have
\beq
(Y_2)_{jk}=q_2^j\delta_{jk}\,,\quad (Y)_{jk}=q^j\delta_{jk}\,,\quad
Y_2^{N}=Y^{2N}=1\,, \label{matY2}
\eeq
and the matrix representation \eqref{CZHQ2} hence describes the 
TBM Hamiltonian $\check{H}_2$ with $q$ given by
\beq
q=e^{\pm i\pi\phi}\,.
\eeq

As to the third representation \eqref{CZHQk}, 
considerig $CZ^\ast$ family for $k\geq 3$ in the same way as above, 
we verify that the matrix \eqref{CZHQk} describes the 
TBM Hamiltonian family $\check{H}_k$ given by \eqref{TBH_k}
\beq
\check{H}_k=i(X^{-1}-X)Y_k+iY_k^{-1}(X^{-1}-X) =\hat{H}(X,Y_k;q_k)\,,
\eeq
where $Y_k$ and $q_k$ are given by
\beq
Y_k=Y^k\,,\quad q_k=q^k=e^{\pm k\pi i\phi}\,.
\eeq

\subsection{generalization of $\check{\mathscr L}_n^\pm$ to other modes $n\not=\pm1$}
\label{rmk:H_kn_model}
\indent

As discussed in Section~\ref{sec:NQCZ}, $X^{\pm2}$ have the effect of increasing or 
decreasing $j$ by 2 ($\Delta j=2$) along the quantum line $X_r$. Reflecting this feature, 
$\hat{H}(X^2,Y;q)$ is deduced to describe the system of which effective interval 
$\Delta j$ is twice that of $\hat{H}(X,Y;q)$. Then denoting another TBM Hamiltonian 
$\hat{H}(X^2,Y;q)$ by $\hat{H}_2$, 
\beq
\hat{H}_2= i(X^{-2}-X^{2})Y + iY^{-1}(X^{-2}-X^{2})\,,
\eeq
we can derive the Schr\"{o}dinger equation
\beq
i(q^{j+2}+q^{-j})\psi_{j+2} - i(q^{-j}+q^{j-2})\psi_{j-2} =E\psi_j
\eeq
by applying $\Delta j=2$ to the original equation \eqref{MBeq}.

The new Hamiltonian $\hat{H}_2$ possesses $U_q(sl_2)$ symmetry: 
\beq
\hat{H}_2=(q_2^{}-q_2^{-1})(\mathcal{E}_+ +\mathcal{E}_-)\,,\quad q_2^{}=q^2\,, \label{Uqsl2_q2}
\eeq
\beq
\mathcal{E}_+=\frac{i}{q_2^{}-q_2^{-1}}(X^{-2}-X^2)Y\,,\quad \mathcal{E}_-
=\frac{i}{q_2^{}-q_2^{-1}}Y^{-1}(X^{-2}-X^2)\,, 
\eeq
\beq
\mathcal{K}=q_2X^{-4}
\eeq
\beq
[\mathcal{E}_+,\mathcal{E}_-]=\frac{\mathcal{K}-\mathcal{K}^{-1}}{q_2^{}-q_2^{-1}}\,,\quad
 \mathcal{K}\mathcal{E}_\pm \mathcal{K}^{-1}=q_2^{\pm2} \mathcal{E}_\pm\,.
\eeq
However it is rather convenient to regard this symmetry as the $n=\pm2$ parts of  $CZ^\ast$ representation family \eqref{CZHQk} when considering the following 
Hamiltonian series  
\beq
\hat{H}_n = i(X^{-n}-X^{n})Y + iY^{-1}(X^{-n}-X^{n})\,,
\eeq
which gives the Schr\"{o}dinger equation with the effective interval $\Delta j=n=2\nu$
\beq
i(q^{j+2\nu}+q^{-j})\psi_{j+2\nu} - i(q^{-j}+q^{j-2\nu})\psi_{j-2\nu} =E\psi_j\,.
\eeq
Namely, as a generalization of Section~\ref{rmk:H_k1_model}, defining 
$\check{H}_{(n,k)}$ in terms of the representation \eqref{CZHQk} of the $CZ^\ast$ family
\begin{align}
\check{H}_{(n,k)}&=(q_k-q_k^{-1})(\check{{\mathscr L}}_n^+ -\check{{\mathscr L}}_{-n}^+)
  +(q_k-q_k^{-1})(\check{{\mathscr L'}}_n^- -\check{{\mathscr L'}}_{-n}^-) \,, \label{Hnk_by_CZ} \\     
&=\hat{H}_n(X,Y_k;q_k)\,, \label{Hnk_by_Hn}
\end{align}
we thus find the connection of the Hamiltonian series $\hat{H}_n$ to 
the $CZ^\ast$ family operators $\check{{\mathscr L}}_n^\pm$, 
which are extended from the $n=\pm1$ modes $\check{{\mathscr L}}_{\pm1}^\pm$. 

Note that the previous cases $\check{H}_1$ and $\check{H}_2$ discussed in Section~\ref{rmk:H_k1_model} belong to the $n=1$ series of $\check{H}_{(n,k)}$
\beq
\check{H}_k=\check{H}_{(1,k)}=\hat{H}(X,Y_k;q_k)\,,\quad \hat{H}_1=\hat{H}\,,\quad k=1,2\,.
\eeq

\subsection{Hamiltonian series with $Y^{\pm k}$ family}
\label{rmk:H22}
\indent

Regarding the first type representation \eqref{CZHQ2}, it may be more convenient to consider the Hamiltonian $\hat{H}(X,Y^2;q)$, instead of the original $\hat{H}(X,Y;q)$. 
In this way, one may anticipate the avoidance of the complicated discussion in the previous subsection and a more direct correspondence between \eqref{CZHQ2} 
and $\hat{H}(X,Y^2;q)$. 

Let us consider the following Hamiltonian family
\beq
\hat{H}_{(n,k)}= i(X^{-n}-X^{n})Y^k + iY^{-k}(X^{-n}-X^{n})\,, \label{Hnk}
\eeq
and first we set $n=k=2$
\beq
\hat{H}_{(2,2)}= i(X^{-2}-X^{2})Y^2 + iY^{-2}(X^{-2}-X^{2})\,.
\eeq
This leads to the following Schr\"{o}dinger equation
\beq
i(q^{2j+4}+q^{-2j})\psi_{j+2} - i(q^{-2j}+q^{2j-4})\psi_{j-2} =E\psi_j\,, \label{Seq22}
\eeq
and it corresponds to a system whose effective interval $\Delta j$ and the size of quantum fluctuation $q$ (see \S\ref{sec:NQCZ}) are twice those of the original 
system \eqref{MBeq}, since \eqref{Seq22} coincides with the equation obtained 
from \eqref{MBeq} by the replacements $\Delta j=1\ra2$ and $q\ra q^2$. It is 
straightforward to verify that the Hamiltonian \eqref{Hnk} describes the quantum 
plane system with the effective interval $n$ and the fluctuation size $q^k$. 

If we introduce the following $\mathscr{H}_n$ operator composed of the generators 
${\mathscr L}_n^\pm$ given in the $CZ^\ast$ representation \eqref{CZHQ2}
\beq
\mathscr{H}_n=(q-q^{-1})({\mathscr L}_n^+ -{\mathscr L}_{-n}^+)
  +(q-q^{-1})(\tilde{{\mathscr L}}_n^- -\tilde{{\mathscr L}}_{-n}^-) \,,   \label{Hn_by_CZ}
\eeq
we find the relation
\beq
\mathscr{H}_2=\hat{H}_{(2,2)}=\hat{H}(X^2,Y_2;q_2)\,.
\eeq
The Hamiltonian $\hat{H}_{(2,2)}$ is also given by a combination of the generators 
of the quantum algebra  $U_q(sl_2)$
\beq
\hat{H}_{(2,2)}=(q_4^{}-q_4^{-1})(\mathcal{E'}_+ +\mathcal{E'}_-)\,,\quad q_4^{}=q^4\,, \label{Uqsl2_q4}
\eeq
\beq
\mathcal{E'}_+=\frac{i}{q_4^{}-q_4^{-1}}(X^{-2}-X^2)Y^2\,,\quad \mathcal{E'}_-=\frac{i}{q_4^{}-q_4^{-1}}Y^{-2}(X^{-2}-X^2)\,, 
\eeq
\beq
\mathcal{K'}=q_4X^{-4}
\eeq
\beq
[\mathcal{E'}_+,\mathcal{E'}_-]=\frac{\mathcal{K'}-\mathcal{K'}^{-1}}{q_4^{}-q_4^{-1}}\,,\quad
 \mathcal{K'}\mathcal{E'}_\pm \mathcal{K'}^{-1}=q_4^{\pm2} \mathcal{E'}_\pm\,.
\eeq
Needless to say, the $CZ^\ast$ representation \eqref{CZHQ2} is again suitable 
to describe the relation between $\mathscr{H}_n$ and the Hamiltonian series 
$\hat{H}_{(n,2)}$
\beq
\mathscr{H}_n=\hat{H}_{(n,2)}=\hat{H}(X^n,Y_2;q_2)\,.   \label{Hn_by_Hn2}
\eeq
We also have its generalization as
\beq
\hat{H}_{(n,k)}=\hat{H}(X^n,Y_k;q_k) \,. \label{Hnk_by_Hnk}
\eeq

As seen above, $\hat{H}_{(n,k)}$ expresses a variety of Hamiltonians designated 
by combinations of effective interval of magnetic lattice $\Delta j$ and quantum 
plane fluctuation $q_k$. There exists a $U_q(sl_2)$ symmetry in the 
$\hat{H}_{(n,k)}$ system for each $n$ and $k$, for example, 
\eqref{Uqsl2_q4} for $\hat{H}_{(2,2)}$,  \eqref{Uqsl2_q2} for $\hat{H}_{(2,1)}$, 
and \eqref{H_by_uqsl2} for $\hat{H}_{(1,1)}$. 
On the other hand, it is possible to express the Hamiltonian series in a unified manner 
such as \eqref{Hnk_by_Hn}, \eqref{Hn_by_Hn2} and \eqref{Hnk_by_Hnk}, 
if we employ one of the closed algebra systems $CZ^\ast$ or $CZ^\ast$ family 
as observed in \eqref{Hnk_by_CZ} and \eqref{Hn_by_CZ}. 

\setcounter{equation}{0}
\section{Conclusions and discussions}
\indent

In this paper we have focused on the relations between $CZ$ algebras and 
the quantum plane picture using algebraic properties of DMT as well as MT. 
The mechanism of generating phase factors is found to be compatible with the 
$\ast$-bracket feature of CZ algebras, and it is very convenient to investigate 
various properties of CZ algebras. As a result, we have clarified some properties 
that could not be obtained from the $q$-harmonic oscillators 
representaion~\eqref{Ln_qho}. 
They are what mechanism determines the phase factor in \eqref{LastL}, 
that it is related to fluctuations on the quantum plane, 
and that there is a certain rule in the method of constructing the CZ generators.
 
Commutative representation is especially important, and we need a specific pair of 
commuting and noncommuting operators. In the MT representation, by introducing 
]the weight of MT and $CZ$ operators, we have presented the definition of 
$\ast$-bracket \eqref{X*X} that can express the three types of $CZ$ algebras, 
$CZ^\pm$ and $CZ^\ast$, in the unified form. The $CZ^\pm$ operator \eqref{Lpm1} 
is a linear combination of commutative $\Op{T}{n}{0}$ and noncommutative 
$\Op{T}{n}{\pm2}$ operators, and the same structure is also found for the DMT 
matrix representation \eqref{QHCZ}. We in fact observed in 
Section~\ref{sec:triviaCZ} that the commutative representations $X^n$ and $Y^n$ 
play the same role as the MT operator $\Op{T}{n}{0}$ as seen in \eqref{TTHQ}. 
The commutative representation shows its significance in connection to 
the $\ast$-brackets and the quantum plane picture. The (non)commutativity of 
the local operator $\Op{T}{n}{0}$  and the nonlocal operator $\Op{T}{n}{\pm2}$ is 
flipped in the $\ast$-bracket commutator as seen in \eqref{T00}. 
The commutative operator carries the essential role of noncommuting $CZ^\pm$ 
algebraic relations, while the noncommuting operator does the role of 
deformed $U(1)$ translational group and thereby determines the weight for 
the $\ast$-bracket \eqref{X*X}. The operators $X^{-n}$ and $X^{-n}Y^{\pm2}$ 
perform the same functions as $\Op{T}{n}{0}$ and $\Op{T}{n}{\pm2}$ 
as mentioned in \eqref{TTHQ}.

The property that commutative operators behave as non-commutative ones 
(and vice versa) matches the phase fluctuation of the quantum plane, and 
by considering the commutative DMT representations \eqref{CZ+Y} and \eqref{CZ-X}, 
we recognize that the $CZ^\pm$ algebras can be described by the $\ast$-ordered products 
\eqref{Yphase} and \eqref{Xphase} in Section~\ref{rmk:q-plane}. In this way, we have 
provided another definition of $\ast$-product by the ordered product that counts the 
number of fluctuations in the positive and negative directions based on the quantum plane 
picture of TBM. The AB phase \eqref{phaseTBM} associated with the movement of particles 
by DMT is interpreted as the fluctuation of the quantum plane \eqref{localq}, and the phase 
factor generated by the successive operations of DMT is then expressed by a certain 
ordered product to reproduce the $\ast$-bracket.

As  a glimpse of the quantum plane picture, we have discussed the relations between 
$CZ^\ast$ algebra and TBM Hamiltonian series in Section~\ref{CZ2TBM}. 
It has been shown that the matrix representation of TBM corresponds to the Wyle representation of $CZ^\ast$, which describes a sequence of  Hamiltonians covering various magnetic lattices. Each Hamiltonian in the sequence is parameterized by two integers $n$ and $k$ which are the power of the Wyle base $X$ and $Y$. 
The parameters $n$ and $k$ correspond to the effective spacing of the magnetic lattice 
and the fluctuation size of quantum plane (power of $q$), respectively. 
All the Hamiltonians can be represented in a unified manner by the $CZ^\ast$ generators 
without introducing additional multiple copies of $U_q(sl_2)$ or $CZ^\ast$ family. 
(Recall that in Section~\ref{sec:preTBM} we have introduced the representation sequence 
of the $CZ^\ast$ family in order to express the TBM Hamiltonian of quantum fluctuation 
size of $k$. The sequence is generated by successively replacing $ q \to q^k.$ However, 
considering a single $CZ^\ast$ algebra with a sequence of Hamiltonians rather than 
the $CZ^\ast$ algebra family is physically easier to understand.) 
In this way, the $CZ^\ast$ algebra may be regarded as a universal algebra to describe 
the Hamiltonian series in accordance to various quantum plane settings of $n$ and $k$. 

The similarities between MT and DMT representations found in this paper may suggest 
a universal property common to various $CZ^\ast$ representations. 
The correspondence between the $q$-differential representation \eqref{Ln_diff} and 
the MT representation \eqref{CZT} may reveal a physical meaning of $q$-differential 
operators in lattice systems with $CZ^\ast$ algebraic structure. The matrix representation 
of TBM Hamiltonian series by Wyle base implies the existence of quantum plane behind 
the physical systems. All these observations are related to the representations of 
$CZ^\ast$ algebra, and we therefore believe that significance of $CZ^\ast$ algebra has 
been increased by this paper. We will be able to clarify unsolved issues and universal 
properties of $CZ^\ast$ algebra from some properties common to multiple representations 
including MT and DMT representations as we have done for the question in \eqref{LastL}. 

\appendix
\setcounter{equation}{0}
\section{quantm group symmetry in TBM}
\label{sec:TBM2CZ}
\indent

In this appendix, we review the quantum algebra symmetry in TBM, that is, 
the TBM Hamiltonian is written by $U_q(sl_2)$ raising-lowering operators~\cite{WZ}. 
We also put a remark on $q$-inversion symmetry of the Hamiltonian. 
In order to see the $U_q(sl_2)$ structure, we impose a periodic condition on the 
Schr\"{o}dinger equation \eqref{TBSeq}, and we then transform \eqref{TBSeq} into a matrix 
form in use of the Wyle base matricis $X$ and $Y$.

Let us choose the factors of hopping terms as in~\cite{HH} so that the gauge invariant 
condition \eqref{phaseTBM} is satisfied
\beq
\theta^x_{m,n}=-(n+m)\pi\phi\,,\quad \theta^y_{m,n}=(m+n+1)\pi\phi\,,
\eeq
and set periodic condition
\beq
\psi_{m,n}=\psi_{m+Q,n+Q}
\eeq
with co-prime integers $P,Q$ and the ratio $\phi$
\beq
\phi=\frac{P}{Q}\,.
\eeq
Under these conditions, \eqref{TBSeq} is invariant under the transformation 
$(m,n)\ra(m+Q,n+Q)$. According to the Bloch theorem, we have
\beq
\psi_{m,n}=e^{imk_x}e^{ink_y}u_{m,n}\,,\quad u_{m,n}=u_{m+Q,n+Q}\,.
\eeq
Using magnetic momenta $k_\pm=\frac{1}{2}(k_x\pm k_y)$ and reflecting 
the exchange symmetry $k_-\lra-k_-$ in $n\lra m$ with
\beq
u_{m,n}=u_{n,m}:=\psi_{n+m}\,,
\eeq
we have
\beq
\psi_{m,n}=e^{ik_+(m+n)}e^{ik_-(m-n)}\psi_{m+n}\,,\quad \psi_{m+n}=\psi_{m+n+2Q}\,.
\eeq
Reducing a degree of freedom with choosing mid band spectrum for 
$k_\pm$~\cite{HH,MB}
\beq
 j=m+n\,,\quad (k_+,k_-)=(\frac{\pi}{2},0)\,,   \label{MBcond}
\eeq
and making $q$ related with $\phi$ by
\beq
q=e^{-i\pi\phi}\,,  \label{q-phi}
\eeq
\eqref{TBSeq} becomes the following equation~\cite{HKW}
\beq
-i(q^{j-1}+q^{-j})\psi_{j-1} + i(q^{-j}+q^{j+1})\psi_{j+1}=E\psi_j\,.  \label{MBeq}
\eeq
This is a component expression of $2Q$-dimensional matrix representation of the Schr\"{o}dinger equation
\beq
\hat{H}\psi=E\psi\,,\quad  \psi=(\psi_1,\psi_2,\cdots,\psi_{2Q})^T  \label{TBMwave}
\eeq
and the Hamiltonian $\hat{H}$ can be written by the matrices $X$ and $Y$ of size $N=2Q$ 
defined in \eqref{XYmat}
\beq
\hat{H}=iY^{-1}(X^{-1}-X)+i(X^{-1}-X)Y\,. \label{MBH}
\eeq
This is known to be written in the $U_q(sl_2)$ generators  $E_\pm$ and $K^{\pm1}$
\beq
E_+=\frac{i}{q-q^{-1}}(X^{-1}-X)Y\,,\quad E_-=\frac{i}{q-q^{-1}}Y^{-1}(X^{-1}-X) \,,
\eeq
\beq
K=q X^{-2}\,,
\eeq
which satisfy the $U_q(sl_2)$ algebraic relations
\beq
[E_+,E_-]=\frac{K-K^{-1}}{q-q^{-1}}\,,\quad KE_\pm K^{-1}=q^{\pm2} E_\pm\,. \label{uqsl2}
\eeq
One can check these relations by using $YX=qXY$, and 
the Hamiltonian reads~\cite{WZ,HH}
\beq
\hat{H}=(q-q^{-1})(E_++E_-)\,.  \label{H_by_uqsl2}
\eeq

Now, let us put some remarks on $q$-inversion symmetry of the Hamiltonian. 
We use the inverted $q=e^{i\pi\phi}$ till the end of this Appendix. 
Considering complex conjugation of \eqref{TBSeq}, we have the Hamiltonian 
\beq
\hat{H}'=Y\hat{H}Y^{-1}=iY(X^{-1}-X)+i(X^{-1}-X)Y^{-1} \,  \label{MBH*}
\eeq	
which looks different from $\hat H$. Of course, $\hat H$ and $\hat{H}'$ describe the 
same physics because $\hat{H}'$ is obtained by the replacement of $q$ in \eqref{MBH} 
with $q^{-1}$ (recall that $Y(q^{-1})=Y^{-1}(q)$). We then notice that for each expression
\beq
\hat{H}^\dagger=\hat{H}\,,\quad \hat{H}^{'\dagger}=\hat{H}'\,.
\eeq

Transformation of any operator $\cal{O}$ to that of $q$-inverted system 
$\cal{O}'$ is similarly given by 
\beq
{\cal O}'=Y{\cal O}Y^{-1}\,,  \label{YOY}
\eeq
and operator relations are identical each other. For example, the $U_q(sl_2)$ 
generators transform as
\beq
E'_\pm=YE_\pm Y^{-1}\,,\quad K'=Y K Y^{-1}
\eeq
and they satisfy the same relations as \eqref{uqsl2}. The inverted Hamiltonian is 
thus given by the identical form to $\hat{H}$
\beq
\hat{H}'=(q-q^{-1})(E'_++E'_-) \,,
\eeq
which is written as the sum of $E'_\pm$ the raising and lowering operators 
of $U_q(sl_2)$.

\setcounter{equation}{0}
\section{matrix representations of DMT}
\label{sec:Txy2XY}
\indent

In Appendix\ref{sec:TBM2CZ}, we have the matrix representation $\hat{H}$ (see \eqref{MBH}) 
of the TBM Hamiltonian $H$ given in \eqref{TBH}. This implies that there is a correspondence 
between the operators $\hat{T}_x$, $\hat{T}_y$ in \eqref{TBH} and the matrices $X$, $Y$ in 
\eqref{MBH}. In this appendix, we clarify the correspondence and then we verify the 
$q$-inversion symmetry \eqref{YOY} and the DMT algebras \eqref{DMTalg} in matrix 
representation. 

In order to see the correspondence, let us consider the $j$-th component of 
matrix actions of $X$ and $Y$ on $\psi$
\beq
(X^{\pm1}\psi)_j=\psi_{j\mp1}\,,\quad (Y^{\pm1}\psi)_j=q^{\pm j} \psi_j \,.\label{XYpsi}
\eeq
These relations mean that $X$ shifts the coordinate $j$ by 1 and $Y$ generates a phase 
factor $q^j$.

{}For all possible products of $X$ and $Y$ appeared in $\hat{H}$ \eqref{MBH} and $\hat{H}'$ \eqref{MBH*}, we have 
\begin{align}
&(X^{\pm1}Y\psi)_j=q^{j\mp1}\psi_{j\mp1}\,,\quad (Y^{-1}X^{\pm1}\psi)_j=q^{-j} \psi_{j\mp1} \,, \\
&(YX^{\pm1}\psi)_j=q^{j}\psi_{j\mp1}\,,\quad (X^{\pm1}Y^{-1}\psi)_j=q^{-j\pm1} \psi_{j\mp1} \,.
\end{align}
On the other hand, acting DMT operators \eqref{TBTxTy} on the wave function \eqref{wavef}, 
for example
\beq
\hat{T}_x\Psi=\sum_{m,n}e^{i\theta^x_{m-1,n}}\psi_{m-1,n}\ket{\psi_{m,n}}
\eeq
and extracting the coefficient of $e^{i\frac{\pi}{2}(m+n)}\ket{\psi_{m,n}}$ 
(since the overall factor $e^{ij\frac{\pi}{2}}$ is excluded when deriving \eqref{MBeq}), 
we obtain
\begin{align}
\hat{T}_x\Psi &\sim e^{-i\frac{\pi}{2}(m+n)}e^{i\theta^x_{m-1,n}}\psi_{m-1,n} \nn \\
&=-i q^{j-1}\psi_{j-1} =-i (XY \psi)_j\,.
\end{align}
Repeating the same process for the rest of DMT operators, we have the following correspondence, namely the matrix representation
\beq
\hat{T}_x\lra -iXY\,,\quad \hat{T}_y\lra -iY^{-1}X\,,\quad 
\hat{T}_x^\dagger\lra iY^{-1}X^{-1}\,,\quad \hat{T}_y^\dagger\lra iX^{-1}Y\,. \label{TbyXY1}
\eeq
Also for the complex conjugate system $(q=e^{i\pi\phi})$ with changing the mid band 
condition $k_+\lra -k_+$ and denoting the DMT operators by $\hat{T}'_x,\hat{T}'_y$ etc., 
we have
\begin{equation}
\hat{T}'_x\lra -iXY^{-1}\,,\quad \hat{T}'_y\lra -iYX\,,\quad 
\hat{T}'_x{}^{\dagger}\lra iYX^{-1}\,,\quad \hat{T}'_y{}^{\dagger}\lra iX^{-1}Y^{-1}\,. \label{TbyXY2}
\end{equation}	
This representation is identical to \eqref{TbyXY1} under the exchange $Y\lra Y^{-1}$ 
and related to \eqref{TbyXY1} by the $q$-inversion \eqref{YOY}
\beq
\hat{T}'_y=Y\hat{T}_x Y^{-1}\,,\quad \hat{T}'_x=Y\hat{T}_y Y^{-1}\,, \quad\mbox{etc.}
\eeq
This transformation is consistent with the Hamiltonian given by \eqref{MBH*}
\beq
\hat{H}'=\hat{T}'_x+\hat{T}'_y+\hat{T}_x^{'\dagger}+\hat{T}_y^{'\dagger}\,.
\eeq

Finally we check the DMT algebras \eqref{DMTalg}. We have the correspondence 
of the exchange \eqref{TxTyCom} and circulation \eqref{TxTyCirc} rules 
using the matrix representation \eqref{TbyXY1}
\beq
\hat{T}_y\hat{T}_x=q^{-2}\hat{T}_x\hat{T}_y\,,\quad
\hat{T}_y^\dagger \hat{T}_x^\dagger \hat{T}_y\hat{T}_x =q^{-2}\,,
\eeq
and these coincide with \eqref{TxTyCom} and \eqref{TxTyCirc} if we remember that 
$q$ is given by \eqref{q-phi}. The fusion rule \eqref{DMTfusion}
\beq
\hat{T}_x\hat{T}_y=q\hat{T}_{x+y}
\eeq
can also be reproduced from the Hermitian conjugate of \eqref{YXfusion}
\beq
\hat{T}_x\hat{T}_y =-X^2 = q \tilde{X}^2
\eeq
and
\beq
\hat{T}_{x+y}=\tilde{X}^2\,.
\eeq

\section*{CRediT author statement} 

\textbf{Naruhiko Aizawa:} Writing-Review \& Editing, Conceptualization, Supervision. 
\textbf{Haru-Tada Sato:} Writing-Original Draft, Conceptualization, Methodology, 
Investigatin, Validation.

\section*{Declaration of competing interest}

The authors declare that we have no known competing financial interests or personal relationships that could have appeared to influence the work reported in this paper.

\section*{Data availability}

No data was used for the research described in the article.

\newcommand{\NP}[1]{{\it Nucl.{}~Phys.} {\bf #1}}
\newcommand{\PL}[1]{{\it Phys.{}~Lett.} {\bf #1}}
\newcommand{\Prep}[1]{{\it Phys.{}~Rep.} {\bf #1}}
\newcommand{\PR}[1]{{\it Phys.{}~Rev.} {\bf #1}}
\newcommand{\PRL}[1]{{\it Phys.{}~Rev.{}~Lett.} {\bf #1}}
\newcommand{\PTP}[1]{{\it Prog.{}~Theor.{}~Phys.} {\bf #1}}
\newcommand{\PTPS}[1]{{\it Prog.{}~Theor.{}~Phys.{}~Suppl.} {\bf #1}}
\newcommand{\MPL}[1]{{\it Mod.{}~Phys.{}~Lett.} {\bf #1}}
\newcommand{\IJMP}[1]{{\it Int.{}~Jour.{}~Mod.{}~Phys.} {\bf #1}}
\newcommand{\JPA}[1]{{\it J.{}~Phys.} {\bf A}:\ Math.~Gen. {\bf #1}~}
\newcommand{\JHEP}[1]{{\it J.{}~High Energy{}~Phys.} {\bf #1}}
\newcommand{\JMP}[1]{{\it J.{}~Math.{}~Phys.} {\bf #1} }
\newcommand{\CMP}[1]{{\it Commun.{}~Math.{}~Phys.} {\bf #1} }
\newcommand{\LMP}[1]{{\it Lett.{}~Math.{}~Phys.} {\bf #1} }
\newcommand{\doi}[2]{\,\href{#1}{#2}\,}  


\end{document}